\documentclass[lettersize]{IEEEtran}
\usepackage{preambles}

\title{Efficient Synthesis for Two-Dimensional Strand Arrays with Row Constraints}

\author{\IEEEauthorblockN{Boaz Moav, Eitan Yaakobi}
  \IEEEauthorblockA{Computer Science, Technion, Haifa, Israel \\
  Email: \{boazmoav, yaakobi\}@cs.technion.ac.il}
\and
\IEEEauthorblockN{Ryan Gabrys}
\IEEEauthorblockA{CalIT2, UCSD, San Diego, California, USA \\
Email: rgabrys@ucsd.edu}}

\author{Boaz Moav,~\IEEEmembership{Member,~IEEE,} Ryan Gabrys,~\IEEEmembership{Member,~IEEE,} Eitan Yaakobi,~\IEEEmembership{Senior Member,~IEEE}
\thanks{B. Moav and E. Yaakobi are with the Henry and Marilyn Taub Faculty of Computer Science, Technion - Israel Institute of Technology, Haifa 3200003, Israel 
(e-mail: \texttt{\{boazmoav,yaakobi\}@cs.technion.ac.il}). 
R. Gabrys is with CalIT2, University of California, San Diego, California, USA,
(e-mail: \texttt{rgabrys@ucsd.edu}).}
\thanks{%
The research was funded by the European Union (ERC, DNAStorage, 101045114 and EIC, DiDAX 101115134). Views and opinions expressed are, however, those of the authors only and do not necessarily reflect those of the European Union or the European Research Council Executive Agency. Neither the European Union nor the granting authority can be held responsible for them.
This work was also supported in part by NSF Grant CCF2212437.}
\thanks{%
This paper was presented in part at ISIT2026.}}

\begin{document}
\maketitle

\begin{abstract}
In large-scale array-based DNA synthesis, optical and chemical coupling between nearby sites can limit simultaneous activations. Motivated by this constraint, we study strands synthesized according to a fixed global synthesis sequence, with at most one strand per row advancing in each cycle.
We focus on the fundamental case of two strands in a single row and analyze the expected completion time of row-constrained synthesis. 
We introduce the \emph{laggard-first (LF) policy}, a simple rule that always advances the strand with fewer synthesized symbols when a conflict arises, and establish that it is asymptotically optimal among online policies without look-ahead. In the binary case, one-symbol look-ahead strictly improves on the no-look-ahead bound. 
We further show that even complete advance knowledge does not eliminate the scheduling loss, as even a globally optimal schedule incurs an unavoidable expected overhead that grows linearly with the strand length.
Finally, we complement these scheduling results with a dynamic programming algorithm for computing an optimal offline synthesis order and a constant-redundancy binary coding scheme that yields a deterministic worst-case synthesis time guarantee. 
\end{abstract}

\section{Introduction}\label{sec:introduction}
DNA-based data storage has emerged as a high-density, long-term alternative to conventional storage, enabled by advances in DNA synthesis and sequencing~\cite{Church2012,Goldman2013}. However, it remains expensive and time-consuming, which motivates continued research. A DNA storage system consists of a synthesizer, a storage medium, and a sequencer, with external coding ensuring reliable reconstruction in the presence of errors. While many parts of this pipeline have been optimized~\cite{Yazdi2017,Lenz2020_2,Rashtchian2017}, synthesis has received comparatively less attention.

Current DNA synthesis machines operate in parallel according to a fixed synthesis sequence. In each cycle, a single nucleotide is appended only to selected strands, as in phosphoramidite chemistry~\cite{Kosuri2014} and photolithographic synthesis~\cite{Antkowiak2020}. Prior work has studied the resulting information density per cycle~\cite{Lenz2020,zrihan2024,Lenz2025}, alternative alphabets~\cite{AbuSini2025}, codes for efficient synthesis~\cite{SchouhamerImmink2024}, and the case in which multiple nucleotides are available per cycle~\cite{Moav2025_2}.

Recent advances in light-directed and enzymatic DNA synthesis indicate that spatial control in dense arrays is limited by optical and chemical coupling between neighboring sites.
Namely, \cite{Antkowiak2020} and \cite{Sack2013} identify optical coupling effects as dominant error sources, motivating increased substrate spacing. This suggests that synthesis requires restricting simultaneous activations within shared optical or chemical regions, which in dense arrays correspond naturally to row-wise constraints. 

\begin{figure}
\centering
\includegraphics[width=\linewidth]{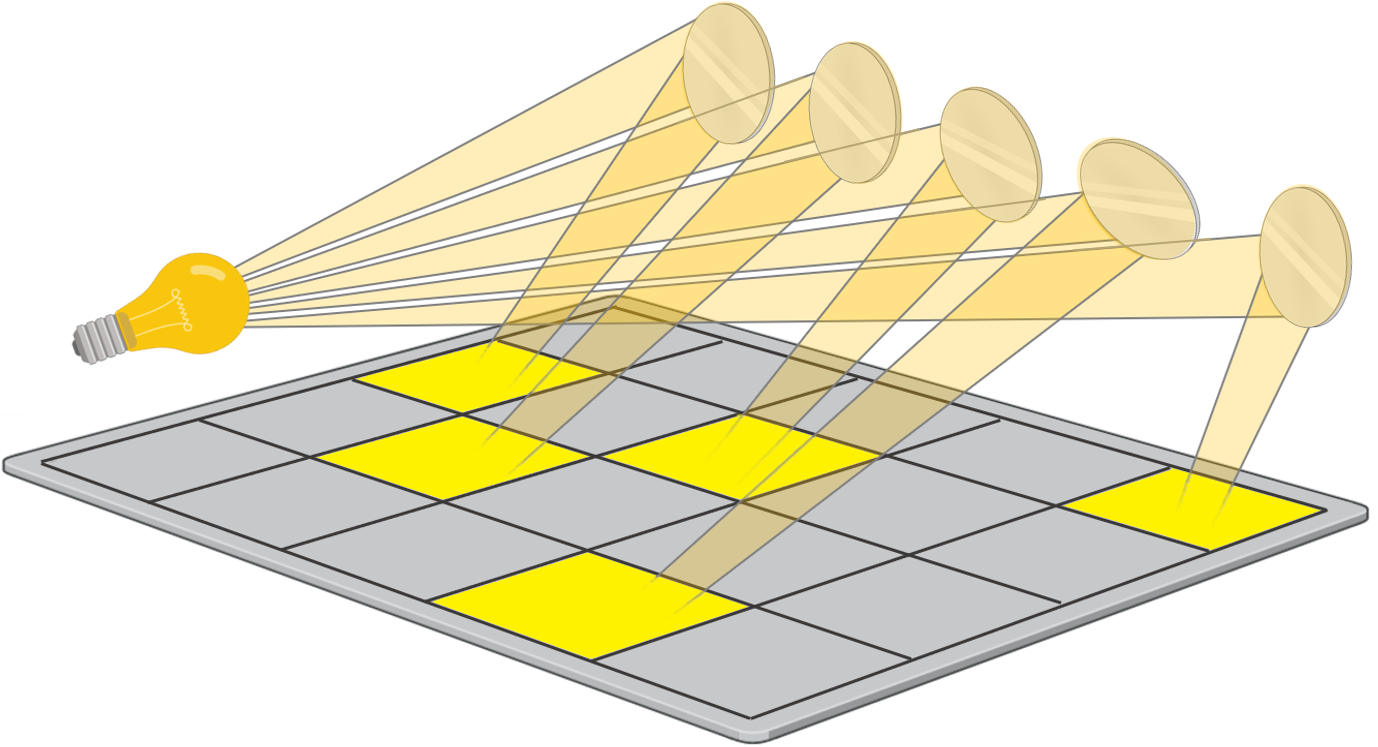}
\caption{Illustration of the array model in light-directed DNA synthesis under neighborhood-wise constraints, e.g., using a separate mirror per row.}
\label{fig:array-model}
\end{figure}

Motivated by these constraints, we model synthesis on an $m \times k$ array, where at most one strand per row may append a symbol per cycle while columns proceed concurrently, as illustrated in \Cref{fig:array-model}. This captures the resource coupling observed in \cite{Sack2013} and enables the analysis of the efficiency limits of spatially constrained parallel synthesis.
Unlike the fully parallel model, where any strand whose next base matches may append this base, the array model enforces row-level coupling, allowing at most one strand per row to append a symbol per cycle. The model was first introduced in~\cite{Moav2025_2}, where the focus was on the offline problem of computing an optimal synthesis sequence for a fixed set of strands, and a dynamic programming algorithm was provided to compute the shortest possible synthesis sequence for a given instance.
However, \cite{Moav2025_2} does not address the average-case behavior or the performance of online policies for random strands.
A preliminary version of this work appeared in~\cite{Moav2026}; the present paper substantially extends it by establishing a strict lower bound on the expected completion time of the optimal offline policy in the binary case, and by introducing an explicit constant-redundancy coding scheme with a worst-case synthesis-time guarantee.

These questions motivate the present work, in which we study expected completion time and its limits under the row-constrained array model. In contrast to prior work~\cite{Moav2025_2}, where the synthesis sequence was optimized for fixed strands, we fix the periodic synthesis sequence and ask how efficiently random strands can be scheduled under the constraint. We focus on the basic but nontrivial case of two strands in a single row, which isolates the scheduling difficulty created by the restriction that at most one strand may advance in each cycle.

We show in \Cref{sec:expected-synth-time} that a simple laggard-first policy, which resolves conflicts by advancing the strand with fewer synthesized symbols, achieves an asymptotic expected completion time of $(q+3)L/2$. We further prove that no online policy without look-ahead can asymptotically improve this leading term. Thus, in the absence of future information, laggard-first is asymptotically optimal.

For the binary case, we show that future information can provably help. A one-symbol look-ahead variant achieves an expected completion time $7L/3+O(1)$, strictly improving upon the no-look-ahead value. In \Cref{sec:lower-bounds-expected-synth-time}, we prove that the trivial lower bound $2L$ is not asymptotically tight. In particular, even an optimal offline policy requires an expected completion time of at least $2.03L-o(L)$. Although the numerical gap is small, this result shows that the overhead above the trivial bound is inherent and not merely a limitation of the policies. 

Finally, we complement the average-case scheduling results with algorithmic and coding results. In \Cref{sec:coding-for-run-max} we show that when the input sequences may be encoded, coding across the two strands can improve the worst-case synthesis time. In the binary case, we construct an explicit constant-redundancy encoding from two length-$(L-9)$ sequences to two length-$L$ sequences that guarantees a synthesis time of at most $8L/3+1$. These results show that scheduling, look-ahead, lower bounds, and coding each capture a different aspect of synthesis efficiency under row-wise spatial constraints.
In \Cref{sec:optimal-order}, we present a dynamic programming algorithm that computes an optimal offline schedule for any fixed pair of input strands.

The remainder of the paper is organized as follows. \Cref{sec:definitions-and-problems} introduces the array model, formally states the three problems studied in this work (\Cref{problem:finding-expected-synthesis-time,problem:finding-optimal-synthesis-order,problem:coding-for-efficien-synthesis}), and surveys prior work. \Cref{sec:expected-synth-time,sec:upper-bounds-binary} addresses \Cref{problem:finding-expected-synthesis-time}, deriving upper bounds on the expected completion time via the laggard-first policy and its one-symbol look-ahead variant. \Cref{sec:lower-bounds-expected-synth-time} establishes matching lower bounds on the expected completion time, including a strict lower bound for the binary case. \Cref{sec:coding-for-run-max} addresses \Cref{problem:coding-for-efficien-synthesis}, presenting an explicit constant-redundancy binary code with a worst-case synthesis time guarantee. \Cref{sec:optimal-order} addresses \Cref{problem:finding-optimal-synthesis-order} via a dynamic programming algorithm for optimal offline scheduling. Proofs deferred from the main text appear in the Appendix.


\Cref{tab:summary_results} summarizes our main results for the row-constrained model, including expected completion times for online policies, worst-case guarantees via coding, and exact offline optimization, together with the corresponding lower bounds.

\begin{table*}[t]
\centering
\caption{Summary of completion time results for synthesizing two sequences of length $L$ over $\Sigma_q$ under the row constraint.}
\label{tab:summary_results}
\renewcommand{\arraystretch}{2.3}
\setlength{\tabcolsep}{4pt}
\begin{tabular}{|m{1.5cm}|m{3.1cm}|m{2.3cm}|m{3.6cm}|m{3.2cm}|m{2cm}|}
\hline
\textbf{Setting} & \textbf{Method / Policy} & \textbf{Metric} & \textbf{Result} & \textbf{Remarks} & \textbf{Proof} \\
\hline
General $q$ & $\bfx$-first 
& $\mathbb{E}[T_{\bfx-\text{first}}(\bfx,\bfy)]$ 
& $\displaystyle \left(\frac{q+3}{2} + \frac{q-1}{q+3}\right)L + O_q(1)$
& Suboptimal 
& \Cref{proposition:x-first-final} \\
\hline
General $q$ & Laggard-first (LF) 
& $\mathbb{E}[T_{\text{LF}}(\bfx,\bfy)]$ 
& $\displaystyle \frac{q+3}{2}L + O_q(1)$ 
& Asymptotically optimal for no look-ahead policies
& \Cref{theorem:LF-synthesis-time} \\
\hline
Binary & LF$_1$ (1-look-ahead) 
& $\mathbb{E}[T_{\text{LF}_1}(\bfx,\bfy)]$ 
& $\displaystyle \frac{7}{3}L + O(1)$ 
& --- 
& \Cref{theorem:binary-one-look-ahead} \\
\hline
Binary & Coding
& $T(\bfx,\bfy)$ 
& $\displaystyle \le \frac{8}{3}L + 1$ 
& $18$ bits of redundancy 
& \Cref{theorem:encoder-binary-synth} \\
\hline
General $q$ & Offline 
& $T_*(\bfx,\bfy)$ 
& DP Algorithm
& Optimal schedule for fixed inputs 
& \Cref{algorithm:min-synth-seq} \\
\hline
$q = 2,3$ & \raggedright Lower bound, Any policy
& $\mathbb{E}[T_*(\bfx,\bfy)]$
& $\displaystyle \ge 2.03L-o(L) > 2L$
& Strictly exceeds trivial $2L$ bound
& \Cref{theorem:lower-bound-for-binary-greater-than-2L} \\
\hline
$q \geq 4$ & \raggedright Lower bound, Any policy
& $\mathbb{E}[T_*(\bfx,\bfy)]$ 
& $\displaystyle \ge \frac{L(q+1)}{2} + O_q(\sqrt{L})$ 
& --- 
& \Cref{lemma:general-lower-bound} \\
\hline
\end{tabular}
\end{table*}

\section{Definitions and Problem Statement}\label{sec:definitions-and-problems}

\subsection{Definitions and Preliminaries}

Let $\Sigma_{q}$ be an alphabet of size $q$. We consider the following model. 
A synthesis device must produce $m \cdot k$ strands arranged in an $m \times k$ array, 
where each cell contains one strand. We say that a strand \emph{advances}, if it appends a 
symbol. In every synthesis cycle, the machine emits a single symbol according to a fixed 
synthesis sequence, and at most one strand per row may advance by incorporating this symbol.
Multiple strands in the same column may advance simultaneously.

In this work, we refer to this as \emph{the array model}. We consider the alternating 
periodic synthesis sequence over $\Sigma_q$, denoted $\bfp_q = (0,1,\ldots,q-1,0,1,\ldots)$,
and study the optimization of the synthesis time. Since different rows evolve independently 
of one another under this constraint, we focus on optimizing the synthesis of one row. 
Next, we define a synthesis schedule for a row in order to find an optimal schedule.
Consider $k$ strands, $\bfx_1, \bfx_2, \ldots, \bfx_k$, each of length $L$ over $\Sigma_q$.
We formalize the synthesis process as a sequence of actions, and we 
refer to the $t$-th element in the sequence as \emph{time slot $t$}.
For a specific row, at time slot $t$, the synthesis machine takes exactly one of the 
following actions:
$\text{X}_1$ - synthesize the next symbol of $\bfx_1$,
$\text{X}_2$ - synthesize the next symbol of $\bfx_2$, etc., and
IDLE - the action of doing nothing.

\begin{definition}
Define a \emph{schedule} as a sequence of actions
$\bfa = (a_1, a_2, \ldots)$, where $a_t \in \{\text{X}_1, \text{X}_2, \ldots, \text{X}_k, \text{IDLE}\}$.
A schedule ends at the first time slot at which all strands have been fully synthesized. 
We refer to $T(\bfx_1, \bfx_2, \ldots, \bfx_k)$ as the \emph{completion time} of the schedule and define a \emph{policy} as a rule that, at each time slot, selects the next action.
\end{definition}

Under a fixed periodic synthesis sequence, action choices may affect the strands' future alignment. The following example illustrates the effect on completion time.

\begin{figure}
\centering
\includegraphics[width=\linewidth]{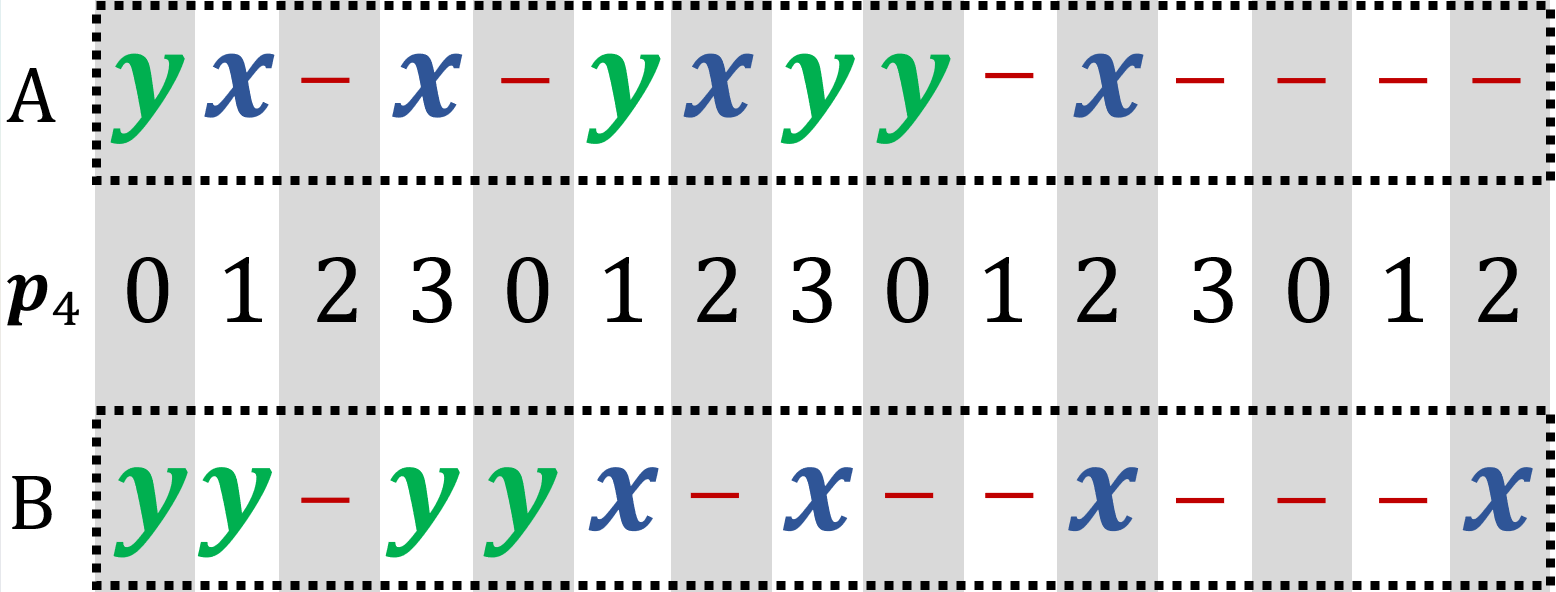}
\caption{Illustration of \Cref{example:ordering-matters}, demonstrating the impact of scheduling under row constraints. The two schedules shown for $\bfx = (1, 3, 2, 2)$ and $\bfy = (0, 1, 3, 0)$ differ in their second time slot.}
\label{fig:ordering-matters}
\end{figure}

\begin{example}\label{example:ordering-matters}
    Let $\bfx = (1, 3, 2, 2)$ and $\bfy = (0, 1, 3, 0)$ over $\Sigma_4 = \{0,1,2,3\}$ and $\bfp_4$ be the alternating periodic synthesis sequence.
    The set of actions is X ($\bfx$ advances), Y ($\bfy$ advances), or $-$ (idle, no sequence advances).
    We present two valid schedules that vary only in how a single tie is resolved, yet they lead to distinct completion times. 
    
    \begin{itemize}
        \item Schedule A (completion time: $11$):
        \begin{flalign*}
            \text{Y}, \text{X}, -, \text{X}, -, \text{Y}, \text{X}, \text{Y}, \text{Y}, -, \text{X}.&&
        \end{flalign*}
        \item Schedule B (completion time: $15$):
        \begin{flalign*}
            \text{Y}, \text{Y}, -, \text{Y}, \text{Y}, \text{X}, -, \text{X}, -, -, \text{X}, -, -, -, \text{X}.&&
        \end{flalign*}
    \end{itemize}
    Both schedules begin identically, synthesizing $y_1$, then the first tie occurs when the current synthesis symbol matches both $y_2$ and $x_1$.
    Schedule A synthesizes $x_1$, while Schedule B synthesizes $y_2$. This single difference misaligns later synthesis opportunities and results in completion times of $11$ and $15$ cycles, respectively.
    This example is also illustrated in \Cref{fig:ordering-matters}. 
\end{example}

\Cref{example:ordering-matters} motivates the following definition.

\begin{definition}\label{definition:optimal-schedule}
An \emph{optimal schedule} is one that minimizes the completion time.
For a policy $\Pi$, let $T_{\Pi}(\bfx_1,\bfx_2,\ldots,\bfx_k)$ be the completion time under the row constraint. Then, define:
$$T_{*}(\bfx_1,\bfx_2,\ldots,\bfx_k) \coloneqq \min_{\Pi} T_{\Pi}(\bfx_1,\bfx_2,\ldots,\bfx_k).$$
\end{definition}
\Cref{definition:optimal-schedule} allows the usage of arbitrary policies, including policies that remain idle even when a strand could advance. Before studying optimal schedules, we first show that such idle actions are never needed for optimality.
\begin{restatable}{lemma}{greedisopt}
\label{lemma:greedy-is-opt}
    There exists an optimal schedule that never remains idle when progress is possible.
\end{restatable}
We provide below the main idea of the proof, and a complete proof is given in the Appendix.
Call a slot a \emph{bad idle} if some strand could advance, but the schedule idles instead. Given any optimal schedule, a bad idle at time $t_0$ can be eliminated by advancing an eligible strand at $t_0$ and replacing the next advance of that strand (at some time $u > t_0$) with an idle. This swap leaves the total number of advances of each strand unchanged, preserves feasibility, and does not alter the completion time. Repeating this exchange eliminates all bad idles, yielding an optimal schedule with none.

Hence, for the purpose of computing an optimal schedule, it is sufficient to restrict attention to schedules that advance some strand whenever progress is possible.
Under this restriction, the only nontrivial choices occur when more than one strand can advance in the same time slot, and the scheduling problem becomes the problem of resolving such conflicts.

\subsection{Problem Statement}\label{subsec:problems}
Let $\bfx \in \Sigma_q^{L}$. Then, $T(\bfx)$ denotes the synthesis time for the case where $k\!=\!1$. This refers to using the periodic sequence without the constraint in the array model, as studied in~\cite{Lenz2020}. As computed in~\cite{Lenz2020}, its expected value is $\mathbb{E}[T(\bfx)]=\frac{L(q+1)}{2}$. This motivates the study of the following question.

\begin{problem}\label[Problem]{problem:finding-expected-synthesis-time}
    Let $\bfx_1,\bfx_2,\ldots,\bfx_k$ be i.i.d. sequences of length $L$, with symbols chosen uniformly at random from $\Sigma_q$.
    Find $\mathbb{E}[T_{*}(\bfx_1,\bfx_2,\ldots,\bfx_k)]$, the expected length of an optimal schedule.
\end{problem}

Next, we discuss the deterministic case.
Note that the minimal possible completion time of $k$ sequences is $kL$ and is obtained by choosing all the $k$ sequences as the alternating periodic sequence of length $L$.
The maximal completion time $qkL$ is obtained by 
$\bfx_1 = \bfx_2 = \dots = \bfx_k = (q-1,\allowbreak q-1,\allowbreak \ldots,q-1).$
Thus, $kL \leq T_{*}(\bfx_1,\bfx_2,\ldots,\bfx_k) \leq qkL$. Consequently, the following problem arises naturally.

\begin{problem}\label[Problem]{problem:finding-optimal-synthesis-order}
    Find the length of an optimal schedule, given $k$ sequences of length $L$.
\end{problem}

A third natural variant arises when the strands  are produced by an encoder.
In this case, the goal is to design the strands so that efficient scheduling is guaranteed.
\begin{problem}\label[Problem]{problem:coding-for-efficien-synthesis}
    Given integers $(k,L)$, and a target synthesis time $T_0$, construct a code $\cC \subseteq (\Sigma_q^L)^k$ such that for every $(\bfx_1,\bfx_2,\ldots,\bfx_k)\in\cC$, we have $T_*(\bfx_1,\bfx_2,\ldots,\bfx_k) \leq T_0$.
\end{problem}

\subsection{Previous Work}
The study of coding and scheduling for efficient DNA synthesis was initiated in~\cite{Lenz2020}, which introduced the problem of encoding information across multiple strands, given a fixed number of synthesis cycles. That work also presented explicit code constructions for the periodic synthesis sequence $\bfp_q$, showing that any strand of length $n$ can be encoded into a sequence of length $n+1$ that is synthesizable within $\left\lfloor\frac{q+1}{2}n\right\rfloor$ cycles.
The authors strengthened these results in~\cite{Lenz2025}.

Several subsequent works have built on these ideas. In \cite{zrihan2024}, the authors proposed new families of codes that achieve high information rates approaching capacity. 
In~\cite{SchouhamerImmink2024}, low-rate synthesis codes were introduced, and in~\cite{SchouhamerImmink2024_2} the authors analyzed the redundancy and asymptotic rate of such constructions for large block lengths. A different synthesis abstraction was considered in~\cite{AbuSini2025}, where shortmers rather than single nucleotides serve as synthesis symbols.

Related generalizations of the synthesis process were considered in~\cite{Moav2025_2}, where a ``complex synthesis'' framework was introduced in which, at each cycle, the synthesizer has access to a subset of $w$ out of the $q$ nucleotides and may append one of them to each strand, subject to per-cycle constraints.

Another line of work studied synthesis under batching and partitioning constraints. In~\cite{Makarychev2022}, the authors considered partitioning the set of sequences into batches and synthesizing each batch separately, deriving bounds on the total number of cycles under optimal partitioning. These bounds were refined in~\cite{Singh2023} using tools from zero-knowledge proof systems.

Synthesis constraints have also been incorporated into error-correcting code design. In \cite{Chrisnata2023}, the authors constructed single-indel-correcting codes whose codewords are restricted to be synthesizable within a fixed number of cycles. This construction was improved in \cite{Nguyen2024}, which also incorporated run-length-limited constraints. In \cite{Lu2024}, codes were developed to correct errors caused by synthesis defects. Finally, \cite{liu2025constrainederrorcorrectingcodesefficient} proposed codes that are synthesizable within $T$ cycles while satisfying  $\ell$-RLL and $\epsilon$-balanced constraints, along with explicit encoding and decoding algorithms and extensions to handle indel errors.

A two-dimensional array model with row-wise synthesis constraints was introduced in~\cite{Moav2025_2}, which formally defined the model studied here and presented a dynamic programming algorithm for computing optimal synthesis schedules.
However, \cite{Moav2025_2} focuses on the offline problem of optimizing the synthesis sequence for a fixed set of strands, and does not address the average-case behavior or the performance of online policies for random strands, which is the focus of the present work.

\section{Upper Bounds on Expected Synthesis Time}\label{sec:expected-synth-time}

We consider the case of two sequences in a single row, i.e., $m=1$, $k=2$. While several results carry over naturally to general $m \times k$ arrays, others require additional work and are deferred to future research. We first obtain upper bounds for arbitrary $q$ by analyzing specific synthesis policies. We then refine these bounds for the binary case, using a one look-ahead policy.

\subsection{Upper Bound for the General Case}\label{subsec:upper-bound-general}
To analyze the expected completion time, we introduce offset variables
that describe which strands can advance in the current synthesis slot.
At the beginning of time slot $t$, let $i_t$ and $j_t$ denote the number of symbols already synthesized from $\bfx$ and $\bfy$, respectively.
Thus, the next symbols to be synthesized are $x_{i_t+1}$ and $y_{j_t+1}$, whenever $i_t,j_t<L$.
Let $r_t$ be the symbol emitted in slot $t$, and define:
\begin{align*}
    a_t &\coloneqq (x_{i_t+1}-r_t)\pmod{q},\\
    b_t &\coloneqq (y_{j_t+1}-r_t)\pmod{q}.
\end{align*}
Hence, $a_t=0$ means that $\bfx$ can advance at slot $t$, and similarly, $b_t=0$ means that $\bfy$ can advance at slot $t$.
The evolution is deterministic between successive visits to $(0,0)$, which is the only state with a policy-dependent decision.
We formalize this below by decomposing the process into intervals between successive visits
to the tie state $(0,0)$, 
and then show different policies for choosing an action at the tie state.

\begin{definition}\label{definition:full-rotation}
A \emph{full rotation} is the interval of slots starting at a visit to $(0,0)$, including
the time slot in which the tie is resolved, and ending just before the next visit
to $(0,0)$. 
Let $T$ be its length.
An \emph{$\bfx$-visit} is a time slot in which $\bfx$ advances, and a \emph{$\bfy$-visit} is a slot in which $\bfy$ advances.
Let $V_X$ and $V_Y$ denote the numbers of $\bfx$-visits and $\bfy$-visits during a full rotation, respectively. 
\end{definition}

For the full rotation analysis below, we consider the corresponding infinite i.i.d. random process, while the boundary effects caused by the finite length $L$ will be handled later.
We begin by showing that each full rotation has a finite exponential moment in \Cref{lemma:T-has-exponential-tail}.

\begin{lemma}\label{lemma:T-has-exponential-tail}
    Let $T$ be the length of one full rotation. Then,
    $$\mathbb{E}[e^{\theta_q T}] < \infty,$$
    for some $\theta_q>0$, depending only on $q$.
\end{lemma}
\begin{IEEEproof}
The process $(a_t,b_t)$ evolves on the finite state space $\{0,1,\ldots,q-1\}^2$.
We show that there exist constants $h_q \in \mathbb{N}$ and $p_q>0$, depending only on $q$, such that from every state, the process reaches $(0,0)$ within $h_q=2q$ slots with probability at least $p_q=1/q$.
Indeed, fix a state $(a,b)$. 
If $a=b$, then both coordinates decrease together until reaching $(0,0)$. 
If $a<b$, then after $a$ slots the process reaches a state of the form $(0,b-a)$. At the next slot, $\bfx$ advances and its new offset is uniform on $\{0,1,\ldots,q-1\}$. With probability $1/q$, this new offset equals $b-a-1$. In that event, the two coordinates become equal and then decrease together to $(0,0)$. 
The case $b<a$ is symmetric.
Therefore, from every state, the probability of returning to $(0,0)$ within at most $2q$ slots is at least $1/q$. Hence, for every $u$,
$$\Pr(T>2q u) \leq (1-1/q)^u.$$
Thus, $T$ has a geometric tail, and therefore has a finite exponential moment.
\end{IEEEproof}

\subsection{The \texorpdfstring{$\bfx$}{x}-First Policy}\label{subsec:x-first-policy}
First, we analyze the case in which each tie is resolved by the decision to let $\bfx$ advance.
Denote it as the \emph{$\bfx\text{-first}$ policy}. 
Under the policy, the initial tie of every full rotation is an $\bfx$-visit.
All other $\bfx$-visits occur at states of the form $(0,b)$, $b\neq 0$, and all $\bfy$-visits occur at states of the form $(a,0)$, $a\neq0$.
It is easier to analyze the behavior of a full rotation using $\bfx$-first.
We will use these results later in the analysis of more general policies.
Next, we provide auxiliary claims for the $\bfx\text{-first}$ policy.

\begin{example}
    Let $q=2$, $\bfp_2=(0,1,\dots)$, $\bfx=(0,1,1)$, and $\bfy=(0,1,1)$. We use the $\bfx$-first policy. At time $t=1$ with $r_1=0$, both next symbols are $0$, so $(a_1,b_1)=(0,0)$ is a tie.
    The policy advances $\bfx$, an $\bfx$-visit, so $\bfx$ moves to its next symbol, $1$, while $\bfy$ remains at $0$.
    Thus, $(a_2,b_2)=(0,1)$, another $\bfx$-visit at $t=2$.
    From this point, the evolution is forced:
    First, only $\bfx$ can advance, yielding the state $(1,0)$ at $t=3$, which is a $\bfy$-visit at $t=3$. Next, only $\bfy$ can advance, yielding $(0,0)$ at $t=4$. Then, the policy advances $\bfx$, completing it. 
    Thus, the states evolve as follows:
    $$ (0,0) \to (0,1) \to (1,0) \to (0,0),$$
    a full rotation of length $3$ (excludes the final state, as it is part of the next full rotation).
    Then, $\bfy$ advances alone, completing at $t=8$.
\end{example}

The next three claims characterize the duration and number of advances in a full rotation under the $\bfx$-first policy. Their complete proofs are deferred to the appendix. Here we provide only the main ideas.

\Cref{claim:E[T]-and-E[X]} gives the expected ratio of $\bfx$-visits, i.e., a connection between $T$ and $V_X$. 
The intuition for \Cref{claim:E[T]-and-E[X]} lies in the fact that we are counting the number of time slots needed to reach $(0,0)$.
Since the next element (in expectation) after an $\bfx$-visit, i.e., state $(0,c)$ ($c \neq 0$), is $( (q+1)/2, c-1 )$, we should expect $(q+1)/2$ time slots (in expectation) between $\bfx$-visits.

\begin{restatable}{claim}{claimETEX}\label[Claim]{claim:E[T]-and-E[X]}
Under the $\bfx\text{-first}$ policy:
$$\mathbb{E}[T] = \frac{q+1}{2}\mathbb{E}[V_X].$$
\end{restatable}
The idea is to track the offset $a_t$ during a full rotation. Since the rotation
begins and ends at $(0,0)$, the increments of $a_t$ telescope to zero. When $a_t=0$, $\bfx$ advances and its new offset is uniform over $\{0,\ldots,q-1\}$, whereas when $a_t>0$, the offset decreases by one.
Applying the optional stopping theorem to the corresponding martingale relates the number of slots to the number of $x$-visits and gives the claimed identity.

\Cref{claim:E[X]-E[Y]} counts the expected advantage of $\bfx$ over $\bfy$ with respect to the number of synthesized bases if we begin a full rotation under the $\bfx$-first policy.

\begin{restatable}{claim}{claimEXEY}\label[Claim]{claim:E[X]-E[Y]}
Under the $\bfx\text{-first}$ policy:
$$\mathbb{E}[V_X] - \mathbb{E}[V_Y] = \frac{2q}{q+1}.$$
\end{restatable}
The idea is to consider the offset difference $d_t=a_t-b_t$ over one full rotation.
As in Claim~\ref{claim:E[T]-and-E[X]}, its increments telescope to zero.
Computing the conditional drift separately at a tie, at an $\bfx$-only advance, and at a $\bfy$-only advance, and then applying optional stopping, expresses the total expected drift in terms of $V_X$ and $V_Y$.

Combining \Cref{claim:E[T]-and-E[X],claim:E[X]-E[Y]}, one obtains two relations among the three quantities $\E[T]$, $\E[V_X]$, and $\E[V_Y]$.
To determine the individual values of $\E[V_X]$ and $\E[V_Y]$, it remains to compute one of them explicitly.
We do this next.

\begin{restatable}{claim}{claimEXEYFinal}
\label[Claim]{claim:E[X]-and-E[Y]-final}
Under the $\bfx\text{-first}$ policy, the following holds:
$$\mathbb{E}[V_X] = \frac{q(q+3)}{2(q+1)},\, \mathbb{E}[V_Y] = \frac{q(q-1)}{2(q+1)}, \, \mathbb{E}[T] = \frac{q(q+3)}{4}.$$ 
\end{restatable}
The idea is first denoting $g(a,b)$ to be the expected number of $\bfx$-visits before the next
return to $(0,0)$ when the current offsets are $(a,b)$. The deterministic evolution when both offsets are positive reduces the system to recurrences for the boundary quantities $A_b \coloneqq g(0,b)$, and $B_a \coloneqq g(a,0)$.
Solving these recurrences gives
$$A_b=\frac{b}{q+1}+\frac{q}{2}, \qquad B_a=-\frac{a}{q+1}+\frac{q}{2}.$$
Substitution gives the value of $\mathbb{E}[V_X]$, and the remaining two identities follow from \Cref{claim:E[T]-and-E[X],claim:E[X]-E[Y]}.

We next use the full-rotation quantities above to evaluate the finite-length performance of the $\bfx$-first policy.
For the $i$-th full rotation, let $(T_i,V_{X,i},V_{Y,i})$ denote its length and the numbers of $\bfx$-visits and $\bfy$-visits, and define
$$X_{\mathrm{tot}}(n) \coloneqq \sum_{i=1}^{n}V_{X,i}, \qquad Y_{\mathrm{tot}}(n) \coloneqq \sum_{i=1}^{n}V_{Y,i}.$$
\begin{restatable}{proposition}{propxfirst}
\label{proposition:x-first-final}
Under the $\bfx\text{-first}$ policy, the following holds:
$$\mathbb{E}[T_{\bfx\text{-first}}(\bfx,\bfy)] = \left(\frac{q+3}{2} + \frac{q-1}{q+3}\right)L + O_q(1).$$
\end{restatable}
The proof idea is as follows. Let
$N_X(L) \coloneqq \min\{n:X_{\mathrm{tot}}(n)\geq L\}$
be the first full rotation during which $\bfx$ is completed, and let $\kappa$ be the number of symbols of $\bfy$ synthesized by that time. A renewal-reward argument gives
$$\mathbb{E}[\kappa] = \frac{\mathbb{E}[V_Y]}{\mathbb{E}[V_X]}L+O_q(1) = \frac{q-1}{q+3}L+O_q(1).$$
Under the $\bfx$-first policy, $\bfx$ evolves as a single independently synthesized strand and therefore requires expected time $(q+1)L/2$. After $\bfx$ is completed, the remaining suffix of $\bfy$ has expected length $L-\kappa$ and is synthesized alone. Combining these two contributions yields the proposition. The full proof can be found in the appendix.

\subsection{The Laggard-First Policy}\label{subsec:lf-policy}
Although the $\bfx$-first policy is suboptimal, its analysis provides the basic full rotation quantities used below.
Specifically, resolving a tie in favor of $\bfy$ gives the same full rotation distribution after swapping the two coordinates. 
We therefore use the $\bfx$-first rotation analysis as a building block for the balanced policy considered next.
Thus, we introduce a policy with strictly better performance than $\bfx$-first and improve the upper bound in \Cref{proposition:x-first-final}.

\begin{definition}\label{definition:LF-policy}
Let $(i_t, j_t)$ be the number of synthesized symbols of $\bfx$ and $\bfy$ (respectively) until time slot $t$.
The \emph{laggard-first} (\emph{LF}) \emph{policy}, denoted as $\text{LF}$, is defined as follows:
\begin{itemize}
    \item If exactly one of $\bfx$ or $\bfy$ can advance, then advance it.
    \item Otherwise, both can advance (``a tie'', i.e., $a_t=b_t=0$):
    \begin{itemize}
        \item If $i_t>j_t$, then advance $\bfy$.
        \item If $i_t\leq j_t$, then advance $\bfx$.
    \end{itemize}
\end{itemize}
\end{definition}

Since LF is a feasible online policy, it provides an explicit upper bound on the optimal expected completion time.
\begin{restatable}{theorem}{lfsynthtime}
\label{theorem:LF-synthesis-time}
Let $\bfx,\bfy\in\Sigma_q^L$ be two random, uniformly distributed sequences. Then,
\begin{align*}
    \mathbb{E}[T_{*}(\bfx,\bfy)] \leq \mathbb{E}[T_{\text{LF}}(\bfx,\bfy)] = \frac{L(q+3)}{2} + O_q(1).
\end{align*}
\end{restatable}
For the binary case, we provide the following intuition for \Cref{theorem:LF-synthesis-time}.
At any time slot, the next synthesis symbol matches the next symbol of at least one strand with a probability of $3/4$, and matches neither with a probability of $1/4$. Thus, an advancement occurs after one synthesis cycle with a probability of $3/4$ and after two synthesis cycles with a probability of $1/4$, yielding $3/4+2\cdot(1/4)=5/4$ synthesis cycles per advancement. Since $2L$ symbols must be synthesized, the expected completion time is $2L \cdot 5/4 = 2.5L$.

We prove \Cref{theorem:LF-synthesis-time} later, after establishing several auxiliary claims. 
The proof separates the analysis into two parts.
First, we estimate the expected time needed to synthesize a total of $2L$ symbols across the two strands, regardless of the policy. 
Then, for the LF policy, we show that the imbalance between the two strands at this time contributes only $O_q(1)$ additional synthesis slots.

For the $i$-th full rotation, under any policy, define
$$Z_i\coloneqq V_{X,i}+V_{Y,i},\quad S_n\coloneqq \sum_{i=1}^n T_i, \quad  A_n\coloneqq \sum_{i=1}^n Z_i.$$
Here, $Z_i$ is the total number of symbols synthesized during the $i$-th full rotation, $S_n$ is the total number of synthesis time slots after $n$ full rotations, and $A_n$ is the total number of synthesized symbols after $n$ full rotations. 
Finally, define the stopping index:
$$N_{2L} \coloneqq \min\{n:A_n\ge 2L\},$$
i.e., the first full rotation by which the cumulative number of synthesized symbols reaches $2L$.
In particular, at this index, at least one of the two strands has accumulated $L$ symbols.

The complete proofs of the following claims are deferred to the appendix.
\begin{restatable}{claim}{claimSNtwoL}
\label[Claim]{claim:S-N-2L-expression}
For any policy that resolves a tie using only past information, the following holds:
    $$
        \mathbb{E}[S_{N_{2L}}] = \mathbb{E}[S_{N_{2L}-1}] = \frac{q+3}{2}L + O_q(1).
    $$
\end{restatable}
The idea is that at the beginning of each full rotation, a no-look-ahead policy may choose whether to advance $\bfx$ or $\bfy$, but the unread symbols remain independent of the past. The two choices produce distributions that differ only by swapping the roles of the strands. Consequently, the pairs $(Z_i,T_i)$ are i.i.d., independently of the particular tie-breaking policy, with $\mathbb{E}[Z_i]=q$, $\mathbb{E}[T_i]=\frac{q(q+3)}{4}$. Applying the renewal-reward estimate at cumulative progress $2L$ gives the result.

The remaining claims in this subsection are specific to the LF policy. 
We keep the same indexed notation for full rotations and define the drift increments
$$\Delta_i \coloneqq V_{X,i}-V_{Y,i},\qquad d_n \coloneqq \sum_{i=1}^n\Delta_i.$$
Thus, $\Delta_i$ is the drift during the $i$-th full rotation, and $d_n$ is the accumulated drift after $n$ full rotations.
In \Cref{claim:imbalance-drift-bound}, we show that the expected accumulated drift at stopping time is finite and depends only on $q$.

\begin{restatable}{claim}{claimdriftimbalance}
\label[Claim]{claim:imbalance-drift-bound}
Under the LF policy, let $N_m \coloneqq \min\{n:A_n\ge m\}$.
Then
$$\sup_{m\ge1}\mathbb{E}[|d_{N_m}|]<\infty.$$
Crucially,
$$\mathbb{E}[|d_{N_{2L}}|]=O_q(1).$$
\end{restatable}
The idea is that under LF, whenever the imbalance is positive, the next tie is resolved in favor of $\bfy$, and whenever it is negative, the next tie is resolved in favor of $\bfx$. Thus, outside a bounded region, the imbalance has a uniform drift toward zero. Together with the exponential-moment bound for the total progress during a full rotation, this gives a geometric drift condition. Decomposing the process into excursions from a bounded region and controlling the maximum imbalance and total progress of each excursion shows that stopping at any cumulative progress level $m$ leaves a uniformly bounded expected imbalance.

Define $N_{*} \coloneqq \min\{n \colon X_{\mathrm{tot}}(n)\ge L \text{ and }Y_{\mathrm{tot}}(n)\ge L\}$, the first full rotation by which both strands have already accumulated at least $L$ synthesized symbols.

\begin{restatable}{claim}{claimextratime}
\label[Claim]{claim:extra-time-is-const}
Under the LF policy, the following holds:
    $$
        \mathbb{E}[S_{N_{*}} - S_{N_{2L}}] = O_q(1).
    $$
\end{restatable}
At the stopping index $N_{2L}$, the total progress is at least $2L$, so the number of symbols still missing from the lagging strand is
controlled by the imbalance $|d_{N_{2L}}|$. 
By \Cref{claim:imbalance-drift-bound}, this deficit has a bounded expectation. Since full rotations have a bounded expected length, completing the remaining deficit requires only $O_q(1)$ additional expected time.

We now restate and prove \Cref{theorem:LF-synthesis-time}.
\lfsynthtime*
\begin{IEEEproof}
    Since LF is a feasible policy, by the definition of $T_{*}(\bfx,\bfy)$, $\mathbb{E}[T_{*}(\bfx,\bfy)] \leq \mathbb{E}[T_{\mathrm{LF}}(\bfx,\bfy)]$.
    It remains to evaluate $\mathbb{E}[T_{\mathrm{LF}}(\bfx,\bfy)]$.
    Let $T_{\mathrm{LF}}(\bfx,\bfy)$ denote the actual completion time of the two finite strands under LF. 
    Before the end of the $(N_{2L}-1)$-th full rotation, fewer than $2L$ symbols have been synthesized in total. 
    Hence, both strands cannot yet be completed, and therefore $S_{N_{2L}-1} \leq T_{\mathrm{LF}}(\bfx,\bfy)$.
    On the other hand, by the end of the $N_{*}$-th full rotation, both strands have
    accumulated at least $L$ symbols.
    Hence, $T_{\mathrm{LF}}(\bfx,\bfy) \leq S_{N_*}$, and thus $S_{N_{2L}-1} \leq T_{\mathrm{LF}}(\bfx,\bfy) \leq S_{N^*}$.
    Taking expectations in the lower bound, by \Cref{claim:S-N-2L-expression},
    $$\mathbb{E}[T_{\mathrm{LF}}(\bfx,\bfy)] \geq \mathbb{E}[S_{N_{2L}-1}] = \frac{q+3}{2}L+O_q(1).$$
    For the upper bound, by \Cref{claim:extra-time-is-const},
    $\mathbb{E}[S_{N_*}-S_{N_{2L}}]=O_q(1),$
    and by \Cref{claim:S-N-2L-expression}, 
    $\mathbb{E}[S_{N_{2L}}]=\frac{q+3}{2}L+O_q(1).$
    Therefore,
    $$\mathbb{E}[T_{\mathrm{LF}}(\bfx,\bfy)] \leq \mathbb{E}[S_{N_*}] = \frac{q+3}{2}L+O_q(1).$$
    Combining the bounds completes the proof.
\end{IEEEproof}

The laggard-first policy is efficient in terms of decision complexity, as it requires no look-ahead. Furthermore, \Cref{theorem:lower-bound-for-optimality-no-lookaheads}, to be presented below, shows that among all online policies without look-ahead, LF is asymptotically optimal in expectation, i.e., no such policy can reduce the leading linear term $(q+3)L/2$ in the expected completion time.

\section{Upper Bounds for the Binary Case}\label{sec:upper-bounds-binary}
We now focus on the binary case $q=2$.
Let $\text{LF}_1$ be the one-symbol look-ahead policy defined as follows.
If exactly one of the two strands can advance in the current slot, then the policy advances that strand.
Suppose instead that both strands can advance, and let the current synthesis symbol be $r\in\{0,1\}$.
If the next symbol of exactly one strand is equal to $1-r$, then $\text{LF}_1$ advances that strand. 
If the look-ahead symbols are equal, then the look-ahead does not distinguish between the two choices, and the policy resolves the tie according to the laggard-first policy, as described in \Cref{definition:LF-policy}.

\subsection{One-Symbol Look-Ahead LF Policy}\label{subsec:one-lookahead-lf-policy}
More formally, we first analyze the corresponding infinite i.i.d. process and later account for the boundary effect caused by finite length $L$.
At the beginning of slot $t$, let $i_t$ and $j_t$ denote the number of symbols already synthesized from $\bfx$ and $\bfy$, respectively, and let $r_t\in\{0,1\}$ be the emitted symbol. 
First, let $a_t$ and $b_t$ be defined as in \Cref{subsec:upper-bound-general}. Then, 
define $c_t \coloneqq (x_{i_t+2}-r_t)\pmod 2$ and $d_t \coloneqq (y_{j_t+2}-r_t)\pmod 2$ for the one-symbol look-ahead rule.
Thus, $a_t=0$ means that $\bfx$ can advance in slot $t$, and similarly, $b_t=0$ means that $\bfy$ can advance.
When $a_t=b_t=0$, the pair $(c_t,d_t)$ determines the one-step look-ahead information. 
If $(c_t,d_t)=(1,0)$, then $\text{LF}_1$ advances $\bfx$; if $(c_t,d_t)=(0,1)$, then $\text{LF}_1$ advances $\bfy$. 
If $c_t=d_t$, then $\text{LF}_1$ resolves the tie by the laggard-first rule; namely, it advances $\bfx$ if $i_t\leq j_t$, and advances $\bfy$ otherwise.
For finite strands, if a look-ahead symbol does not exist as the strand is left with a final symbol, we treat that strand as having no favorable look-ahead match.

\begin{theorem}
\label{theorem:binary-one-look-ahead}
Let $\bfx,\bfy\in\Sigma_2^L$ be binary random, uniformly distributed sequences. Then,
$$
    \mathbb{E}[T_{*}(\bfx,\bfy)] \leq \mathbb{E}[T_{\text{LF}_1}(\bfx,\bfy)] = \frac{7}{3}L + O(1).
$$
\end{theorem}
\begin{IEEEproof}
Since $\text{LF}_1$ is a feasible policy, the first inequality follows from the definition of $T_{*}(\bfx,\bfy)$.
It remains to analyze $\mathbb{E}[T_{\mathrm{LF}_1}(\bfx,\bfy)]$.

We begin with the infinite i.i.d. process. Let
$$S_t \coloneqq (a_t,b_t,c_t,d_t)\in\{0,1\}^4,$$
be the state at time slot $t$, and define the total-progress reward 
$$Z(S_t) \coloneqq \begin{cases}
1, & \text{if at least one strand advances in slot }t,\\
0, & \text{otherwise}.
\end{cases}$$
Equivalently, $Z(S_t)=0$ only when $a_t=b_t=1$. Let
$$G_t \coloneqq \sum_{\ell=0}^{t-1} Z(S_\ell),$$
be the total number of synthesized symbols by time $t$.
Since at most one strand advances in each slot, $G_t$ increases by at most $1$ per slot.
Lastly, define $\tau_m \coloneqq \min\{t: G_t\geq m\}.$

The actual process $S_t$ under $\mathrm{LF}_1$ is not Markov on $\{0,1\}^4$, because when $a_t=b_t=0$ and $c_t=d_t$, the decision depends on the imbalance $i_t-j_t$.
Nevertheless, the total-progress process can be analyzed by a finite-state auxiliary chain. Define $\overline{\mathrm{LF}}_1$ to be the policy that agrees with $\mathrm{LF}_1$ except in the two unresolved look-ahead states $(0,0,0,0)$ and $(0,0,1,1)$, where it always advances $\bfx$.
Under $\overline{\mathrm{LF}}_1$, the process $S_t$ is a Markov chain on $\{0,1\}^4$.

We now justify why this auxiliary rule preserves the total-progress process. 
Let $\sigma(a,b,c,d)=(b,a,d,c)$ be the operation that swaps the roles of $\bfx$ and $\bfy$. The reward $Z$ is invariant under this swap.
Moreover, in the only states where $\mathrm{LF}_1$ and $\overline{\mathrm{LF}}_1$ may differ, choosing $\bfx$ or choosing $\bfy$ produces next-state distributions that are swapped by $\sigma$.
Since $Z$ is constant on each swap, the process $(G_t)_{t\geq 0}$, and hence the stopping
times $\tau_m$, have the same distribution under $\mathrm{LF}_1$ as under the auxiliary chain $\overline{\mathrm{LF}}_1$, up to the initial transient states handled below.

Although $P$ in \Cref{figure:P-pi-for-one-look-ahead} is written as a $16\times 16$ matrix, it is not irreducible on all of \(\{0,1\}^4\).
With the state ordering by binary representation, states $12$ and $14$ are transient: they leave this set in one step and are never re-entered. Let
$$\mathcal C \coloneqq \{0,1,\ldots,15\} \setminus \{12,14\}$$ be the closed recurrent class of states.

A direct computation from the transition matrix $P$ in \Cref{figure:P-pi-for-one-look-ahead} gives $\pi$ as the stationary distribution of the closed recurrent class of the auxiliary chain $\overline{\mathrm{LF}}_1$. 
Then,
$$R \coloneqq \sum_{s\in\cC} \pi_s Z(s) = \frac{6}{7}.$$
Thus, $R$ is the asymptotic number of synthesized symbols per synthesis slot.
We next convert this total-progress rate into an expected hitting time.

Fix a state $s^\star\in\cC$ with $Z(s^\star)=1$.
For the cycle argument, w.l.o.g. start the auxiliary chain from $s^\star$, since the expected hitting time of $s^\star$ is $O(1)$, and define
$$\eta_0 \coloneqq 0,\quad \eta_{k+1}:=\min\{t>\eta_k:S_t=s^\star\}.$$
For $k\geq 1$, let
$$ C_k \coloneqq \eta_k-\eta_{k-1},\quad U_k \coloneqq \sum_{t=\eta_{k-1}}^{\eta_k-1} Z(S_t).$$
For each $s\in\cC$, define
$$\gamma_s \coloneqq \mathbb E_{s^\star}\left[ \sum_{t=0}^{\eta_1-1}\mathbf 1\{S_t=s\} \right].$$
$\gamma=(\gamma_s)_{s\in\mathcal C}$ is an invariant measure by \cite[Theorem 1.7.5]{Norris1998}, i.e., $\gamma P=\gamma$ for all $j\in\mathcal C$.
Since the chain is finite and irreducible, the invariant distribution is unique. 
Therefore, the normalized invariant measure $\gamma/\sum_{s\in\mathcal C}\gamma_s$ must equal $\pi$.
Moreover,
$$\sum_{s\in\mathcal C}\gamma_s = \mathbb E_{s^\star}\left[\sum_{t=\eta_0}^{\eta_{1}-1}\sum_{s\in\mathcal C}\bfone\{S_t=s\} \right] = \mathbb E[C_1].$$
Thus,
$$\pi_s=\frac{\gamma_s}{\mathbb E[C_1]}.$$
Now we compute the expected reward in one cycle:
\begin{align*}
\mathbb E[U_1] &=\mathbb E_{s^\star}\left[ \sum_{t=\eta_0}^{\eta_{1}-1} Z(S_t) \right] \\
&= \sum_{s\in\mathcal C} Z(s) \mathbb E_{s^\star}\left[ \sum_{t=\eta_0}^{\eta_{1}-1}\bfone\{S_t=s\} \right] = \sum_{s\in\mathcal C} Z(s)\gamma_s.
\end{align*}
Dividing by $\mathbb E[C_1]$, we obtain
$$\frac{\mathbb E[U_1]}{\mathbb E[C_1]} = \sum_{s\in\mathcal C} Z(s)\frac{\gamma_s}{\mathbb E[C_1]} = \sum_{s\in\mathcal C}\pi_s Z(s).$$
Hence,
$$\frac{\mathbb{E}[U_1]}{\mathbb{E}[C_1]} = \sum_{s\in\mathcal C}\pi_s Z(s) = R = \frac{6}{7}.$$

Since the chain restricted to $\mathcal C$ is finite and irreducible, the return time $C_k$ has a finite second moment. Also, $0\leq U_k\leq C_k$.
Hence, the pairs $(U_k,C_k)$ are i.i.d. with finite second moments, and $U_k \geq 1$ because $Z(s^\star)=1$.
Moreover, for $N_m \coloneqq \min\{n \colon \sum_{k=1}^{n} U_k \geq m\}$, it holds that
$$\sum_{k=1}^{N_m-1}C_k \leq \tau_m  \leq \sum_{k=1}^{N_m} C_k.$$
Thus, applying \Cref{lemma:expectation-of-overshoot} with $R_i=U_i$ and $W_i=C_i$, we obtain
$$\mathbb{E}[\tau_m]=\frac{m}{R}+O(1)=\frac{7}{6}m+O(1).$$
Taking $m=2L$, this gives
$$\mathbb{E}[\tau_{2L}]=\frac{7}{3}L+O(1).$$

The estimate above was derived for the auxiliary policy $\overline{\mathrm{LF}}_1$.
However, as we explained in the beginning of the proof, the same holds for $\mathrm{LF}_1$, and the difference lies only in the imbalance between the two strands at $\tau_{2L}$.
Let $D_t \coloneqq i_t-j_t$ be the imbalance between the two strands according to $\overline{\mathrm{LF}}_1$.
By \Cref{lemma:LF1-bounded-imbalance},
$\mathbb{E}\bigl[|D_{\tau_{2L}}|\bigr]=O(1)$.

At time $\tau_{2L}$, exactly $2L$ symbols have been synthesized in the infinite process, since $Z(s)\in\{0,1\}$.
It remains to pass from the infinite process to the finite length-$L$ strands.
Indeed, the finite and infinite processes can be coupled so that they agree until one of the two original length-$L$ prefixes is completed. 
After this time, any discrepancy is caused only by advances that the infinite process assigns to a strand after its first $L$ symbols have already been synthesized.
At time $\tau_{2L}$, the two infinite counts satisfy $i_{\tau_{2L}}+j_{\tau_{2L}}=2L$, and therefore the number of missing symbols in the lagging length-$L$ prefix is at most $\left|D_{\tau_{2L}}\right|/{2}.$
Since each additional binary symbol appears within at most two synthesis slots, the boundary discrepancy is bounded by a constant multiple of $|D_{\tau_{2L}}|+1$.
Hence, using $\mathbb E[|D_{\tau_{2L}}|]=O(1)$, we conclude that
$$\mathbb{E}[T_{\mathrm{LF}_1}(\bfx,\bfy)] = \mathbb{E}[\tau_{2L}]+O(1) = \frac{7}{3}L+O(1).$$

\end{IEEEproof}

\begin{figure*}[t]
    \begin{align*}
    \mathbf{P} =\left(\begin{array}{cccccccccccccccc}
        0 & 0 & 0 & 0 & 0 & 0 & 0 & 0 & 0 & 0 & 0 & 0 & 0 & \frac{1}{2} & 0 & \frac{1}{2} \\
        0 & 0 & 0 & 0 & 0 & 0 & 0 & 0 & 0 & 0 & \frac{1}{2} & \frac{1}{2} & 0 & 0 & 0 & 0 \\
        0 & 0 & 0 & 0 & 0 & \frac{1}{2} & 0 & \frac{1}{2} & 0 & 0 & 0 & 0 & 0 & 0 & 0 & 0 \\
        0 & 0 & 0 & 0 & \frac{1}{2} & 0 & \frac{1}{2} & 0 & 0 & 0 & 0 & 0 & 0 & 0 & 0 & 0 \\
        0 & 0 & 0 & 0 & 0 & 0 & 0 & 0 & 0 & \frac{1}{2} & 0 & \frac{1}{2} & 0 & 0 & 0 & 0 \\
        0 & 0 & 0 & 0 & 0 & 0 & 0 & 0 & \frac{1}{2} & 0 & \frac{1}{2} & 0 & 0 & 0 & 0 & 0 \\
        0 & \frac{1}{2} & 0 & \frac{1}{2} & 0 & 0 & 0 & 0 & 0 & 0 & 0 & 0 & 0 & 0 & 0 & 0 \\
        \frac{1}{2} & 0 & \frac{1}{2} & 0 & 0 & 0 & 0 & 0 & 0 & 0 & 0 & 0 & 0 & 0 & 0 & 0 \\
        0 & 0 & 0 & 0 & 0 & 0 & \frac{1}{2} & \frac{1}{2} & 0 & 0 & 0 & 0 & 0 & 0 & 0 & 0 \\
        0 & 0 & \frac{1}{2} & \frac{1}{2} & 0 & 0 & 0 & 0 & 0 & 0 & 0 & 0 & 0 & 0 & 0 & 0 \\
        0 & 0 & 0 & 0 & \frac{1}{2} & \frac{1}{2} & 0 & 0 & 0 & 0 & 0 & 0 & 0 & 0 & 0 & 0 \\
        \frac{1}{2} & \frac{1}{2} & 0 & 0 & 0 & 0 & 0 & 0 & 0 & 0 & 0 & 0 & 0 & 0 & 0 & 0 \\
        0 & 0 & 0 & 1 & 0 & 0 & 0 & 0 & 0 & 0 & 0 & 0 & 0 & 0 & 0 & 0 \\
        0 & 0 & 1 & 0 & 0 & 0 & 0 & 0 & 0 & 0 & 0 & 0 & 0 & 0 & 0 & 0 \\
        0 & 1 & 0 & 0 & 0 & 0 & 0 & 0 & 0 & 0 & 0 & 0 & 0 & 0 & 0 & 0 \\
        1 & 0 & 0 & 0 & 0 & 0 & 0 & 0 & 0 & 0 & 0 & 0 & 0 & 0 & 0 & 0
        \end{array}\right), \qquad
    \pi = \begin{pmatrix}
        1/7 \\
        1/21 \\
        11/84 \\
        1/28 \\
        1/18 \\
        13/126 \\
        11/252 \\
        23/252 \\
        13/252 \\
        1/36 \\
        19/252 \\
        13/252 \\
        0 \\
        1/14 \\
        0 \\
        1/14
    \end{pmatrix}
    \end{align*}
    \caption{$\mathbf{P}$ and $\pi$ for the proof of \Cref{theorem:binary-one-look-ahead}.}
    \label{figure:P-pi-for-one-look-ahead}
\end{figure*}

\subsection{Connection to the Chvátal-Sankoff Constants}\label{subsec:chavtal-sankoff-const}
We interpret constrained synthesis combinatorially by relating the optimal completion time to classical longest-common-subsequence (LCS) problems.
Denote an interleaving of $\bfx$ and $\bfy$ as $\bfz \in \{0,1\}^{2L}$, namely a merge preserving the internal order of each strand. 
Synthesizing $\bfz$ alone under the periodic sequence, as in~\cite{Lenz2020}, is equivalent to synthesizing $\bfx$ and $\bfy$ under the row constraint, since $\bfz$ specifies which strand advances at each cycle.
Let $\rho_{\bfz}$ be the number of run transitions in $\bfz$, i.e., the number of indices $i$ such that $z_{i+1} \neq z_{i}$. 
Since changing symbols costs one synthesis cycle and repeating a symbol costs two synthesis cycles, the synthesis time for $\bfz$ is 
\begin{equation}
    1 + \bfone_{\{z_1=1\}} + \rho_{\bfz} + 2(2L - 1 - \rho_{\bfz}) = 4L - 1 + \bfone_{\{z_1=1\}} - \rho_{\bfz}, \label{eq:synth-time-as-runs}
\end{equation}
since the initial symbol costs one cycle if $z_1=0$ and two if $z_1=1$.
Thus, minimizing the completion time is equivalent to maximizing the number of run transitions in $\bfz$.
Also, the optimal completion time satisfies 
$T_{*}(\bfx,\bfy) = \min_{\bfz \in Z(\bfx,\bfy)} T(\bfz),$
where $Z(\bfx,\bfy)$ denotes the set of all interleavings of $\bfx$ and $\bfy$. Equivalently, $T_{*}(\bfx,\bfy)$ is achieved by the interleaving $\bfz$ that maximizes the number of run transitions.

Observe that a lower bound on the number of run transitions in $\bfz$, i.e., the number of adjacent indices for which $z_i \neq z_{i+1}$, can be generated from matched symbols between $\bfx$ and the bitwise complement $\overline{\bfy}$.
Specifically, the minimal number of run transitions is lower-bounded by $2 \cdot \text{LCS}(\bfx,\overline{\bfy})$, and one can construct a $\bfz$ that has $2 \cdot \text{LCS}(\bfx,\overline{\bfy})$ symbols belonging to the longest common subsequence and $2 \cdot (L-|\text{LCS}(\bfx,\overline{\bfy})|)$ symbols that do not, which is an upper bound on the optimal $\bfz$. 
Thus, the following result follows.
\begin{align*}
	T_{*}(\bfx,\bfy) &\leq 2 \cdot \text{LCS}(\bfx,\overline{\bfy}) + 2\cdot 2 \cdot (L-|\text{LCS}(\bfx,\overline{\bfy})|)\\
    &=2L \left( \frac{|\text{LCS}(\bfx,\overline{\bfy})|}{L} + 2\cdot \frac{L-|\text{LCS}(\bfx,\overline{\bfy})|}{L} \right).
\end{align*}

\begin{example}
    Let $\bfx \!=\! (0,1,1,0)$, $\bfy \!=\! (0,1,0,0)$. Matching $x_1$, $x_2$, $x_3$ with $\overline{y}_2$, $\overline{y}_3$, $\overline{y}_4$ yields $\text{LCS}(\bfx,\overline{\bfy}) = (0,1,1)$.
    We denote these symbols as the matched symbols.
    Thus, one can construct the following $\bfz = (y_1, y_2, x_1, x_2, y_3, x_3, y_4, x_4) = (0,1,0,1,0,1,0,0)$. 
    Note that each time one encounters an unmatched symbol (such as in the last position), one can spend at most two cycles. Each time one goes from an unmatched symbol to a pair of matched symbols, we can pick the order in which to present the matched symbols so that this transition costs at most one cycle. Here, we chose $y_2$, since it was not equal to $y_1$. 
    The bound gives $T_{*}(\bfx,\bfy) \leq 10$, but we succeeded in constructing $\bfz$ such that $T(\bfz) = 9$, using the choice to begin the matched symbols with $y_2$ over $x_1$.
\end{example}

The study on the expected LCS length of two random binary sequences was introduced by Chvátal and Sankoff in \cite{Chvtal1975}. Using the improved lower bound for the binary Chvátal-Sankoff constant found in \cite{Lueker2009}, $\mathbb{E}[\text{LCS}(\bfx,\bfy)] \geq 0.788L$, this yields
$T_{*}(\bfx,\bfy) \lesssim 2L \left( 0.788 + 2 \cdot 0.212 \right) = 2.424 L.$
While this bound is weaker than the $7L/3$ bound achieved in \Cref{theorem:binary-one-look-ahead}, it is conceptually appealing. Our future research plan is to explore it for larger alphabets as well.

\section{Lower Bounds on Expected Synthesis Time}\label{sec:lower-bounds-expected-synth-time}

Having established upper bounds on the expected synthesis time, we now derive lower bounds that hold for 
arbitrary scheduling policies, regardless of their computational power or access to future information. 
The trivial lower bound is $2L$, since the schedule must perform a total of $2L$ advances. The main result of this
section is that this bound is not asymptotically attainable for any fixed alphabet size $q\geq2$, and for
$q > 3$ is at least $L(q+1)/2 + O_q(\sqrt{L})$. Thus, even an optimal offline policy incurs
an unavoidable linear scheduling loss for every fixed alphabet size.

\subsection{General Case}
This simple lower bound is obtained by observing that both strands must individually be synthesizable.
\begin{lemma}\label{lemma:general-lower-bound}
Let $q>2$. Then, for two random, uniformly distributed sequences $\bfx,\bfy\in\Sigma_q^L$, the following holds.
$$\mathbb{E}[T_{*}(\bfx,\bfy)] \geq \mathbb{E}[\max(T(\bfx),T(\bfy))] = \frac{L(q+1)}{2} + O_q(\sqrt{L}).$$
\end{lemma}
\begin{IEEEproof}
Denote the random variables $X=T(\bfx)$ and $Y=T(\bfy)$. 
Since $\max(X, Y) = \frac{X+Y+|X-Y|}{2}$ and $X \sim Y$,
\begin{align*}
\mathbb{E}[\max(X, Y)] &= \frac{1}{2} \mathbb{E}[X+Y+|X-Y|] \\
&= \mathbb{E}[X] + \frac{1}{2} \mathbb{E}[|X-Y|] \\
&= L(q+1)/2 + \mathbb{E}[|X-Y|]/2.
\end{align*}
To analyze $\mathbb{E}[|X-Y|]$, we denote
$Z \coloneqq X-Y = \sum_{i=1}^{L} (U_i-V_i)$, where $U_i$ and $V_i$ are the random variables of a single symbol, i.e., i.i.d. uniform on $\{1, 2, \ldots, q\}$.
Thus, if $D_i=U_i-V_i$, then $\mathbb{E}[D_i]=0$ and since $U_i$ and $V_i$ are independent, $\Var(D_i) = \Var(U_i) + \Var(V_i) = 2 \cdot \Var(U_i) = \frac{q^2-1}{6}$.
Then, by the central limit theorem, $Z$ converges in distribution to $\cN(0,\sigma^2)$, where $\sigma^2 = L \cdot \frac{q^2-1}{6}$.
Since $Z$ converges in distribution to a centered Gaussian random variable with variance $\sigma^2$, one obtains asymptotically,
\begin{align*}
    \mathbb{E}[|X-Y|] &= \mathbb{E}[|Z|] \approx \sigma \cdot \sqrt{\frac{2}{\pi}}
    = \sqrt{\frac{2}{\pi}} \cdot \sqrt{\frac{L(q^2-1)}{6}} \\
    &= \sqrt{\frac{L(q^2-1)}{3\pi}}
\end{align*}
Finally,
\begin{align*}
    &\mathbb{E}[\max(X, Y)] = \frac{L(q+1)}{2} + \frac{1}{2} \mathbb{E}[|X-Y|] \\
    &\approx \frac{L(q\!+\!1)}{2} \!+\! \sqrt{\frac{L(q^2\!-\!1)}{12\pi}} \!+\! o_q(\sqrt{L}) = \frac{L(q\!+\!1)}{2} + O_q(\sqrt{L}).
\end{align*}
\end{IEEEproof}

For $q\geq4$, \Cref{lemma:general-lower-bound} already gives a linear-order separation from $2L$. When $q=3$, however, its leading term is exactly $2L$. The additional term is only of order $\sqrt{L}$ and therefore does not rule out convergence of the normalized expected synthesis time to $2$. The following corollary relates the ternary minimum synthesis time to its binary counterpart.

\begin{restatable}{corollary}{ternaryproj}\label{corollary:ternary-binary-projection}
For $q\geq2$, let $\tau_q(n)$ denote the expected minimum synthesis time of two independent uniformly distributed words in $\Sigma_q^n$. Set
$$
 m_L\coloneqq\left\lfloor\frac{2L}{3}-L^{2/3}\right\rfloor.
$$
If $\tau_2(m_L)\geq 2.03m_L-o(m_L)$ as $m_L\to\infty$, then
$$
\tau_3(L) \geq 2.03L-o(L).
$$
Consequently,
$$
\liminf_{L\to\infty}\frac{\tau_3(L)}{L} \geq 2.03>2.
$$
\end{restatable}
The idea is to project each ternary word onto a binary subsequence by deleting all occurrences of the symbol $2$.
With high probability, both projected words contain at least $m_L$ symbols, while deleting the $2$-slots from any ternary schedule produces a valid binary schedule for their length-$m_L$ prefixes.
This gives a $3/2$ scaling between the ternary and binary completion times, up to an additive constant, and hence transfers the binary lower bound to the ternary case. The full proof appears in the appendix.

Our objective is to show that, for every fixed $q\geq2$, the expected minimum synthesis time exceeds the trivial lower bound $2L$ by a term that is linear in $L$. 
For $q=3$, \Cref{corollary:ternary-binary-projection} shows that any linear separation proven for the binary case also holds in the ternary case. It therefore remains only to establish a linear separation when $q=2$. This is the purpose of \Cref{theorem:lower-bound-for-binary-greater-than-2L}, which we state and prove in the next subsection.

\subsection{Binary Case}
We now turn to the binary case, which is the key remaining case. Let $\bfx,\bfy\in\{0,1\}^L$ be independent, uniformly distributed sequences. Since a schedule must perform $2L$ advances and at most one strand can advance in each cycle, every policy satisfies $\mathbb{E}[T_{*}(\bfx,\bfy)]\geq2L$. We prove that this counting bound is not asymptotically attainable. Even an optimal offline policy incurs an expected number of additional synthesis cycles that grows linearly with $L$. The proof translates this scheduling loss into an upper bound on the maximum number of run transitions achievable by interleaving the two strands.
The run-transition accounting in \Cref{eq:synth-time-as-runs} gives
\begin{equation}
    T_{*}(\bfx,\bfy) \geq 4L - 1 - R_L(\bfx,\bfy), \label{eq:opt-synth-time-as-runs}
\end{equation}
where 
$$R_L(\bfx,\bfy) = \max_{{\bfz} \in Z(\bfx,\bfy)} \rho_{\bfz}.$$
This is the maximum number of run transitions achievable by interleaving $\bfx$ and $\bfy$.
Thus, to lower bound the synthesis time, it is enough to upper bound $R_L(\bfx,\bfy)$.
\begin{definition}\label{definition:normalized-expected-run-count}
    Let $\bfx,\bfy\in\{0,1\}^L$ be two random, uniformly distributed sequences. Define
    $$\gamma(L) \coloneqq \frac{1}{2L}\mathbb{E}[R_L(\bfx,\bfy)],$$
    which is the expected maximum number of run transitions, normalized by $2L$.
\end{definition}

A direct result from \Cref{definition:normalized-expected-run-count} is presented below in \Cref{lemma:limit-of-gamma-exists}.

\begin{lemma}\label{lemma:limit-of-gamma-exists}
The following limit exists.
$$\lim_{L\to\infty}\gamma(L).$$
\end{lemma}
\begin{IEEEproof}
Let $a_L \coloneqq \mathbb E[R_L(\mathbf x,\mathbf y)]$.
We show that the sequence $(a_L)_{L\geq 1}$ is superadditive.
Fix $m,n \geq 1$, and let $\bfx_1,\bfy_1\in\{0,1\}^m$ and $\bfx_2,\bfy_2\in\{0,1\}^n$.
Let $\bfz_1 \in Z(\bfx_1, \bfy_1)$ and $\bfz_2 \in Z(\bfx_2, \bfy_2)$ be interleavings satisfying $\rho_{\bfz_1} = R_m(\bfx_1, \bfy_1)$ and $\rho_{\bfz_2} = R_n(\bfx_2, \bfy_2)$.

Then, the concatenation $\bfz_1\bfz_2$ is an interleaving of $\bfx_1\bfx_2$ and $\bfy_1\bfy_2$, namely, $\bfz_1\bfz_2 \in Z(\bfx_1\bfx_2, \bfy_1\bfy_2)$.
Moreover, by the definition of $\rho_z$,
$$\rho_{\bfz_1\bfz_2} = \rho_{\bfz_1}+\rho_{\bfz_2} +\bfone\{(\bfz_1)_{2m} \neq (\bfz_2)_1\} \geq \rho_{\bfz_1}+\rho_{\bfz_2}.$$
Therefore, 
$$R_{m+n}(\bfx_1\bfx_2,\bfy_1\bfy_2) \geq R_m(\bfx_1,\bfy_1)+R_n(\bfx_2,\bfy_2).$$
Therefore, applying the deterministic inequality above and taking expectations over randomly chosen sequences gives 
\begin{align*}
    a_{m+n} &= \mathbb E[R_{m+n}(\bfx_1\bfx_2,\bfy_1\bfy_2)] \\
    &\geq \mathbb E[R_m(\bfx_1,\bfy_1)] + \mathbb E[R_n(\bfx_2,\bfy_2)] = a_m+a_n.
\end{align*}
Thus, $(a_L)_{L\geq 1}$ is superadditive.
By Fekete's lemma~\cite{Fekete1923}, 
$$\lim_{L\to\infty}\frac{a_L}{L} = \sup_{L\geq 1}\frac{a_L}{L},$$
exists, and $\lim_{L\to\infty}\gamma(L)$ exists as well.
\end{IEEEproof}

Following \Cref{lemma:limit-of-gamma-exists}, we now state the main result of this subsection.
\Cref{theorem:lower-bound-for-binary-greater-than-2L} shows that the normalized optimal run-transition density stays strictly below $1$ asymptotically.
Equivalently, for independent uniformly random input strands, even an optimal schedule incurs a positive linear expected number of IDLE slots.
This gives a strict asymptotic improvement over the trivial $2L$ lower bound for binary sequences.

\begin{restatable}{theorem}{lowerboundETforbinary}\label{theorem:lower-bound-for-binary-greater-than-2L}
There exists a constant $\delta>0$ such that
$$\lim_{L \to \infty} \gamma(L) \leq 1-\delta.$$
Moreover, one may take $\delta=0.015$, yielding
$$\mathbb{E}[T_{*}(\bfx,\bfy)] \geq (2+2\delta)L-o(L) = 2.03L-o(L).$$
\end{restatable}

We first introduce the terminology used in the proof.
For $\bfz\in\{0,1\}^{2L}$, let
$$
D(\bfz)\coloneqq
\left\{(\bfx,\bfy)\in\{0,1\}^{L}\times\{0,1\}^{L}
\mathrel{\big|}\bfz\in Z(\bfx,\bfy)\right\}.
$$
We say an index $k\in\{1,\ldots,2L-3\}$ is \emph{alternating} for $\bfz$ if
$$
z_k\neq z_{k+1},\qquad z_{k+1}\neq z_{k+2},\qquad
z_{k+2}\neq z_{k+3}.
$$
And because the alphabet is binary, the corresponding factor has the form $abab$ with $a\neq b$.

To bound the cardinality of $D(\bfz)$, we represent each pair in this set by a coloring of $\bfz$.
A \emph{witness} for a pair $(\bfx,\bfy)\in D(\bfz)$ is a coloring
$$\boldsymbol{\omega}=\omega_1\cdots\omega_{2L}\in\{\mathrm{R},\mathrm{B}\}^{2L},$$
in which the symbols of $\bfz$ in the red positions, read from left to right, form $\bfx$, while those in the blue positions form $\bfy$.
For example, if $\bfz=0111$, then $\mathrm{RRBB}$ witnesses the pair $(01,11)$. E.g., $\mathrm{RBRB}$ also represents the pair $(01,11)$.
Every pair in $D(\bfz)$ has at least one witness, but distinct witnesses may represent the same pair.
Thus, counting witnesses can overcount $|D(\bfz)|$.

One type of overcounting occurs on the alternating word $\bfz=0101$.
The distinct colorings $\mathrm{RRBB}$ and $\mathrm{BBRR}$ both witness the pair $(01,01)$.
Thus, counting both colorings counts the same element of $D(0101)$ twice.
More generally, this equivalence holds for every alternating length-four factor.
If such a factor is colored $\mathrm{RRBB}$, then its first copy of $ab$ is assigned to the red subsequence and its second copy to the blue subsequence.
Recoloring it $\mathrm{BBRR}$ assigns the same word $ab$ to each subsequence and therefore leaves the witnessed pair unchanged.
To eliminate this source of overcounting, we use $\mathrm{BBRR}$ as the canonical representative of these two equivalent colorings and say that a witness $\boldsymbol{\omega}$ is \emph{reduced} if for every alternating index $k$,
$\omega_k\omega_{k+1}\omega_{k+2}\omega_{k+3}\neq\mathrm{RRBB}$.

This restriction is imposed only at alternating indices, where the equivalence
$\mathrm{RRBB}\sim\mathrm{BBRR}$ holds.
As shown below in \Cref{lemma:reduced-witness-bound}, every pair in $D(\bfz)$ retains at least one reduced witness.
Therefore, the number of reduced witnesses still upper-bounds $|D(\bfz)|$, while avoiding this source of overcounting.

We now outline the proof for \Cref{theorem:lower-bound-for-binary-greater-than-2L}.
For a fixed interleaving $\bfz$, the witness-reduction lemma, \Cref{lemma:reduced-witness-bound}, formalizes the local replacement above and bounds $|D(\bfz)|$ by the number of reduced witnesses.
A pattern-avoidance estimate then bounds the number of reduced witnesses for every near-alternating word $\bfz$.
Combining this estimate with an entropy bound on the number of near-alternating words and a union bound gives an exponentially small probability of achieving too many run transitions.

We begin by formalizing the witness-reduction step.
\begin{lemma}\label{lemma:reduced-witness-bound}
Every pair in $D(\bfz)$ has a reduced witness.
Consequently, $|D(\bfz)|$ is at most the number of reduced witnesses for $\bfz$.
\end{lemma}
\begin{IEEEproof}
Fix $(\bfx,\bfy)\in D(\bfz)$ and any witness that produces it.
If an alternating factor $abab$ is colored $\mathrm{RRBB}$, change those four colors to $\mathrm{BBRR}$.
Before the change, the red subsequence receives $ab$ from the first two positions and the blue subsequence receives $ab$ from the last two positions.
After the change, each subsequence still receives $ab$ in the same relative location and the new coloring witnesses the same pair.

Repeat this replacement whenever $\mathrm{RRBB}$ occurs at an alternating index.
Each replacement increases the sum of the indices of the red positions, $S(\boldsymbol{\omega}) = \sum_{i \colon \omega_i = \mathrm{R}} i$, by $4$.
This sum is bounded because there are only $2L$ positions, thus the procedure terminates with a reduced witness of the original pair.
Choose one such reduced witness for each pair in $D(\bfz)$.
Different pairs receive different witnesses, since a witness determines exactly one pair, proving the claimed injection.
\end{IEEEproof}

It remains to count reduced witnesses.
Let $\lambda$ be the real root greater than $1$ of
$ t^3-t^2-t-1=0$. So $\lambda \approx 1.83928675$.
Set
$$
\kappa\coloneqq\frac{\lambda}{2-4\lambda^{-3}}
\approx 1.35525976.
$$
The next estimate follows from a standard count of words that avoid $\mathrm{RRBB}$.
Its proof appears 
in the appendix, immediately after the proof of \Cref{corollary:ternary-binary-projection}.

\begin{restatable}{lemma}{nearalternatinginterleavingbound}
\label{lemma:fixed-near-alternating-interleaving-bound}
Fix $0\leq\beta<1$.
If $\bfz\in\{0,1\}^{2L}$ satisfies
$\rho_{\bfz}\geq2L(1-\beta)$, then
$$
|D(\bfz)|\leq\lambda^{2L}\kappa^{2\beta L}.
$$
\end{restatable}

We next combine this estimate with an entropy bound to prove the main result.
For the reader's convenience, we restate \Cref{theorem:lower-bound-for-binary-greater-than-2L} before giving its proof.

\lowerboundETforbinary*
\begin{IEEEproof}
For $0<\delta<1/2$, let
$$ N_L(\delta)\coloneqq \left\{\bfz\in\{0,1\}^{2L}\mathrel{\big|} \rho_{\bfz}\geq2L(1-\delta)\right\} $$
and
$$ A_L(\delta)\coloneqq \left\{R_L(\bfx,\bfy)\geq2L(1-\delta)\right\}. $$
In words, $N_L(\delta)$ contains the length-$2L$ binary words that are nearly alternating, with at least $2L(1-\delta)$ run transitions.
The event $A_L(\delta)$ occurs when $\bfx$ and $\bfy$ admit at least one interleaving from $N_L(\delta)$.
For every $\bfz\in N_L(\delta)$,
\Cref{lemma:fixed-near-alternating-interleaving-bound} gives $|D(\bfz)|\leq\lambda^{2L}\kappa^{2\delta L}$.

A binary word $\bfz\in\{0,1\}^{2L}$ is uniquely determined by its first symbol $z_1$ and the set $E(\bfz)\coloneqq\{i\in[2L-1] \colon z_i=z_{i+1}\}$ of positions at which it does not change. 
For every $\bfz\in N_L(\delta)$,
$$\left|E(\bfz)\right| \leq 2L-1-2L(1-\delta) =2\delta L-1 \leq \lfloor 2\delta L\rfloor.$$
There are two choices for $z_1$, and, for each $j$, there are $\binom{2L-1}{j}$ choices for a set of $j$ equality positions. Consequently,
$$\left|N_L(\delta)\right| \leq 2\sum_{j=0}^{\lfloor 2\delta L\rfloor} \binom{2L-1}{j}.$$
Letting $k=\lfloor2\delta L\rfloor$ and using the standard entropy estimate $\binom{n}{m} \leq 2^{n h(m/n)}$,
we obtain
\begin{align*}
|N_L(\delta)|
&\le
2 \sum_{j=0}^{\lfloor 2\delta L \rfloor}
\binom{2L-1}{j} \\
&\le
2^{2L h(\delta) + O(\log L)},
\end{align*}
where \(h(\cdot)\) denotes the binary entropy function.

There are $4^L$ equally likely input pairs.
The event $A_L(\delta)$ can occur only if the pair belongs to $D(\bfz)$ for some $\bfz\in N_L(\delta)$.
The union bound therefore gives
\begin{align*}
\Pr(A_L(\delta))
&\leq4^{-L}\sum_{\bfz\in N_L(\delta)}|D(\bfz)|\\
&\leq4^{-L}\left|N_L(\delta)\right| \max_{\bfz\in N_L(\delta)}\left|D(\bfz)\right|\\
&\leq 4^{-L} 2^{2Lh(\delta)+O(\log L)} \lambda^{2L} \kappa^{2\delta L}.\\
&\leq2^{c(\delta)L+O(\log L)},
\end{align*}
where
$c(\delta)\coloneqq 2\log_2\lambda+2\delta\log_2\kappa+2h(\delta)-2$.
Next, for every pair $(\bfx,\bfy)$,
$$ R_L(\bfx,\bfy) \leq 2L(1-\delta)+2L\bfone_{A_L(\delta)}.$$
Taking expectations gives
$$
\mathbb{E}[R_L(\bfx,\bfy)]
\leq2L(1-\delta)+2L\Pr(A_L(\delta)).
$$
If we set $\delta=0.015$, then $c(0.015)\approx-0.0038286<0$, and the probability bound gives
$$\Pr(A_L(0.015)) \leq2^{c(0.015)L+O(\log L)} =e^{-\Omega(L)} = o(1).$$
Thus, substituting $\delta=0.015$ into the expectation bound yields
\begin{align*}
\mathbb{E}[R_L(\bfx,\bfy)]
&\leq2L(1-0.015)+2L\Pr(A_L(0.015))\\
&=1.97L+o(1).
\end{align*}
Dividing by $2L$ and using \Cref{lemma:limit-of-gamma-exists} gives
$$
\lim_{L\to\infty}\gamma(L)\leq0.985=1-0.015.
$$
Using \Cref{eq:opt-synth-time-as-runs}, the same expectation bound also gives
$$\mathbb{E}[T_*(\bfx,\bfy)] \geq 2.03L-o(L).$$
More generally, $c(0)<0<c(1/2)$ and
$$
c'(\delta)=2\log_2\left(\frac{\kappa(1-\delta)}{\delta}\right)>0
\qquad (0<\delta<1/2).
$$
Thus, $c$ has a unique positive zero at
$\approx 0.0152962764$, and the coefficient in the lower bound can be taken arbitrarily close to $\approx 2.0305925529$.
\end{IEEEproof}

\subsection{No Look-ahead Lower Bound}
Denote the class of policies that resolve a tie using only past information (no look-ahead) as $\cP_{0}$.
We provide a lower bound demonstrating the optimality of \Cref{theorem:LF-synthesis-time}.
\begin{theorem}
\label{theorem:lower-bound-for-optimality-no-lookaheads}
Let $\bfx,\bfy\in\Sigma_q^L$ be random, uniformly distributed sequences. 
Then, for every $\Pi \in \cP_{0}$, it holds that,
$$\mathbb{E}[T_{\Pi}(\bfx,\bfy)] \geq \frac{L(q+3)}{2} + O_q(1).$$
\end{theorem}
\begin{IEEEproof}
    Fix an arbitrary $\Pi \in \cP_{0}$ and consider the states $(a_t,b_t)$, where 
    \begin{align*}   a_t &\coloneqq (x_{i_t+1}-r_t)\pmod{q},\\
    b_t &\coloneqq (y_{j_t+1}-r_t)\pmod{q}. \end{align*}
    in which $\bfx_{j}$ is the next symbol in $\bfx$, $\bfy_{j}$ is the next symbol in $\bfy$, and $r \in \Sigma_q$ is the symbol to synthesize in slot $t$.
    Using the same definitions of \Cref{theorem:LF-synthesis-time} for a full rotation, note that between two visits to $(0,0)$, the evolution of $(a_t,b_t)$ is deterministic. Whenever both coordinates are positive, both are decremented. Whenever exactly one is zero, that strand necessarily advances and is reset uniformly on ${0,\ldots,q-1}$.
    Thus, the behavior inside a full rotation depends only on the underlying randomness, not on how ties at $(0,0)$ are resolved. 
    The only effect of the policy $\Pi$ is to decide, at the first slot of each full rotation, whether that full rotation starts with an $\bfx$-advance or a $\bfy$-advance.
    Let the random variables $X_i$, $Y_i$, $T_i$, $Z_i$, $A_n$, $S_n$, $X_{\text{tot}}(n)$, and $Y_{\text{tot}}(n)$ be defined as in \Cref{theorem:LF-synthesis-time}.
    Then, by \Cref{claim:E[X]-and-E[Y]-final},
    $$
        \mathbb{E}[Z_i] = q, \quad 
        \mathbb{E}[T_i] = \frac{q(q+3)}{4}.
    $$
    Thus, define the following stopping index.
    $$N_{2L} \coloneqq \min \left\{ n \colon A_n \geq 2L \right\}.$$
    By \Cref{claim:S-N-2L-expression},
    $$
        \mathbb{E}[S_{N_{2L}}] = \mathbb{E}[S_{N_{2L}-1}] \frac{q+3}{2}L + O_q(1).
    $$
    Define the true stopping time for policy $A$ as
    $$N_{*} \coloneqq \min \left\{ n \colon X_{\text{tot}}(n) \geq L \text{ and } Y_{\text{tot}}(n) \geq L\right\}.$$
    Then, $N_{*} \geq N_{2L}$ since the event $\{ X_{\text{tot}}(n) \geq L \text{ and } Y_{\text{tot}}(n) \geq L \}$ implies $\{A_n \geq 2L\}$.
    Thus, we conclude that
    $$
        \mathbb{E}[T_{\Pi}(\bfx,\bfy)] \geq \mathbb{E}[S_{N_{*}-1}] \geq \mathbb{E}[S_{N_{2L}-1}].
    $$
\end{IEEEproof}

\section{A Constant-Redundancy Coding Scheme for Run Maximization}\label{sec:coding-for-run-max}

In this section, we present a binary coding scheme for a pair of length-$L$ sequences that induces an interleaving with many runs and, therefore, guarantees a bounded synthesis time under the row-constrained model.
The construction is based on grouping symbols into triples, controlling repeated symbols within each triple, and then controlling repeated symbols across adjacent triples. The scheme uses only constant redundancy of $18$ bits.
We assume throughout that $L$ is divisible by $3$. The general case can be handled by appending at most two padding symbols, which changes the final bound by at most an additive constant.

Let $\bfx, \bfy \in \{0,1\}^{L-9}$. We present an encoding that maps $(\bfx,\bfy)$ to a pair of sequences $(\bfx',\bfy') \in \{0,1\}^L \times \{0,1\}^L$ such that there exists a feasible synthesis schedule for $(\bfx',\bfy')$ of length at most $\frac{8L}{3}+1$.
To obtain this result, we will construct a $2L$-length sequence, which is an interleaving $\bfz$ of $\bfx'$ and $\bfy'$, such that $\rho_{\bfz}$, the number of runs in $\bfz$, is maximized, thus applying the conclusion from \Cref{subsec:chavtal-sankoff-const}.
Define:
$$ M = \frac{2L - 18}{3}.$$

We construct $M$ disjoint blocks, each containing three bits.
Each block is denoted as $B_i=((a_i,b_i),c_i)$, $i\in [M]$, where $(a_i,b_i)$ are two consecutive bits from one strand (the order must be preserved), and $c_i$ is a bit from the other strand.
Blocks alternate between two forms: odd-indexed triples take the next two unused consecutive bits from $\bfx$ and the next unused bit from $\bfy$, while even-indexed triples take the next two unused consecutive bits from $\bfy$ and the next unused bit from $\bfx$.
In the sequel, we will specify how each block is locally interleaved, and by concatenating these local interleavings, we obtain a global interleaving of all symbols.
Thus, the sequence of blocks implicitly defines an interleaving of $\bfx$ and $\bfy$ that respects the order constraints in both strands. 

\begin{example}\label{example:split-to-M-blocks}
    Let $\bfx=(\textcolor{olive}{0},\textcolor{olive}{0},\textcolor{olive}{0})$ and $\bfy=(\textcolor{magenta}{0},\textcolor{magenta}{1},\textcolor{magenta}{0})$. Thus, $M=2$, i.e., two blocks of triples, as follows:
    \begin{align*}
        B_1 = \{(\textcolor{olive}{0},\textcolor{olive}{0}),\textcolor{magenta}{0}\},\,\, B_2 = \{(\textcolor{magenta}{1},\textcolor{magenta}{0}),\textcolor{olive}{0}\}.
    \end{align*}
    Thus, the first block will always interleave as $(0,0,0)$, while the second block can be interleaved as either $(0,1,0)$, or $(1,0,0)$, but not as $(0,0,1)$, which breaks the ordering. These interleavings are the local interleavings.
    Concatenating the locally interleaved blocks yields an interleaving of $\bfx$ and $\bfy$. For example, $(\textcolor{olive}{0},\textcolor{olive}{0},\textcolor{magenta}{0}, \textcolor{olive}{0}, \textcolor{magenta}{1},\textcolor{magenta}{0})$.
\end{example}

There are exactly $8$ possible block types.
We define a canonical local interleaving rule for a single triple block.
For a block $B=((a,b),c)$, define its greedy interleaving $g(B) \in \{0,1\}^3$ by
$$ g(\{(0,0),1\})=g(\{(0,1),0\})=g(\{(1,0),0\})=(0,1,0),$$
and
$$ g(\{(1,1),0\})=g(\{(1,0),1\})=g(\{(0,1),1\})=(1,0,1).$$
Lastly, for the two homogeneous blocks $\{(0,0),0\}$ and $\{(1,1),1\}$, the interleaving contains two repeated adjacencies.
Thus, for every block except the two homogeneous blocks, there exists a sound interleaving with no repeated adjacent symbols, and we assume the local interleavings are performed only in this way.

The two homogeneous blocks contribute exactly two repeated adjacencies.
Therefore, one wishes to reduce the number of homogeneous blocks and thus increase the number of runs.
Let $\cB$ be the set of $8$ block types, and consider a permutation $\sigma \colon \cB \to \cB$ such that $\sigma$ is applied to each of the $M$ blocks.

\begin{lemma}\label{lemma:good-permutation-existance-for-non-homogeneous}
    There exists a permutation $\sigma$ such that, after applying $\sigma$ to all blocks, the number of homogeneous blocks is at most $\lfloor M / 4 \rfloor$.
\end{lemma}
\begin{IEEEproof}
    Let $\cH \subset \cB$ be the set consisting of the two homogeneous block types. For each permutation $\sigma$ of $\cB$, let $N_\sigma$ be the number of positions $i \in [M]$ such that $\sigma(B_i) \in \cH$.
    We average $N_\sigma$ over all $8!$ permutations. Fix a position $i$. 
    As $\sigma$ ranges over all permutations of $\cB$, the image $\sigma(B_i)$ takes each value in $\cB$ exactly $7!$ times.
    Since $|\cH|=2$, it follows that $\sigma(B_i)$ is homogeneous for exactly $2 \cdot 7!$ permutations, and the fraction of permutations for which $\sigma(B_i)$ is homogeneous is:
    $$\frac{2 \cdot 7!}{8!} = \frac{2}{8} = \frac{1}{4}.$$
    Summing over all $M$ positions, the average value of $N_\sigma$ is therefore $M/4$. Consequently, there exists at least one permutation $\sigma$ such that
    $N_{\sigma} \leq M/4$.
    Since $N_{\sigma}$ is an integer, this yields
    $$N_{\sigma} \leq \left\lfloor\frac{M}{4}\right\rfloor,$$
    which completes the proof.
\end{IEEEproof}

Denote the total number of repeated adjacencies inside blocks as $R_{\text{in}}$.
From \Cref{lemma:good-permutation-existance-for-non-homogeneous}, there exists a permutation $\sigma$, such that concatenating $(\sigma(B_1), \sigma(B_2), \ldots, \sigma(B_M))$ yields the following:
$$R_{\text{in}} \leq 2\left\lfloor\frac{M}{4}\right\rfloor.$$

Next, we handle repetitions between consecutive blocks after concatenation.
We say that two consecutive blocks $(B_i, B_{i+1})$ are \emph{unfriendly} if their greedy interleavings create a repeated adjacency at the boundary, i.e., the last bit of $B_i$ and the first bit of $B_{i+1}$. Otherwise, the blocks are \emph{friendly}.
Define the complement of a block:
$$\overline{((a,b),c)} = ((1-a,1-b),1-c),$$
and note that if $(B_i, B_{i+1})$ is unfriendly, then $(B_i, \overline{B}_{i+1})$ is friendly.
Denote by $\phi$ the function that flips all even-indexed blocks.

\begin{lemma}\label{lemma:flipping-for-friendly-blocks}
    By either keeping the blocks unchanged or applying $\phi$, the number of unfriendly boundaries is at most:
    $$\left\lfloor \frac{M-1}{2} \right\rfloor.$$
\end{lemma}
\begin{IEEEproof}
    There are $M-1$ boundaries. Let $U$ be the number of unfriendly boundaries. After flipping even blocks, the number becomes $(M-1)-U$. One of these two choices is at most half.
\end{IEEEproof}

Denote the total number of repeated adjacencies in the boundaries as $R_{\text{boundary}}$.
By either applying $\phi$ or not, there exists a concatenation of the blocks such that:
$$R_{\text{boundary}} \leq \left\lfloor \frac{M-1}{2} \right\rfloor.$$
Combining these results and denoting $R$ as the total number of repetitions in the encoded payload, which is an interleaving of $\bfx$ and $\bfy$ after applying $\sigma$ and possibly $\phi$:
$$R \leq  2\left\lfloor\frac{M}{4}\right\rfloor + \left\lfloor \frac{M-1}{2} \right\rfloor.$$
Since $M=\frac{2L-18}{3}$ is even,
$R \leq M-1 = 2L/3 - 7$.

The encoded payload has a length of $2L-18$; therefore, the number of runs, $\rho$, equals $(2L-18) - 1 - R$, which results in a lower bound on the number of runs in the encoded payload:
$$\rho \geq (2L-18) - 1 - \left( \frac{2L}{3} - 7 \right) = \frac{4L}{3} - 12$$
Thus, we conclude, using \Cref{eq:synth-time-as-runs}, that an upper bound on the synthesis time of the encoded payload is
\begin{align}
    T_{\text{pld}} \leq 2((2L-18)-1) - \left(\frac{4L}{3} - 12\right) = \frac{8L}{3} - 26. \label{eq:T-payload}
\end{align}

To store the permutation $\sigma$, one needs $\left\lceil\log_2(8!)\right\rceil = 16$ redundancy bits.
Additionally, one needs one bit to store whether $\phi$ was used or not.
Finally, to ensure that these $17$ bits are also synthesized efficiently, we encode them using the encoder from~\cite{Lenz2020}, with one additional bit of redundancy to ensure  that \begin{align}
    T_{\text{red}} \leq \frac{3 \cdot 18}{2}=27.  \label{eq:T-redundancy}
\end{align}

Thus, we conclude with the following theorem.
\begin{theorem}\label{theorem:encoder-binary-synth}
    For any $\bfx, \bfy \in \{0,1\}^{L-9}$, where $L$ is divisible by 3, there exists an explicit encoding $(\bfx, \bfy) \to (\bfx^{\prime}, \bfy^{\prime}) \in \{0,1\}^{L} \times \{0,1\}^{L}$, such that:
    $$T_{*}(\bfx^{\prime}, \bfy^{\prime}) \leq \frac{8L}{3} + 1.$$
    That is, for large enough $L$, the synthesis time is $\approx 2.66L$.
\end{theorem}
\begin{IEEEproof}
    Given $(\bfx, \bfy) \in \{0,1\}^{L-9}$, the encoder constructs $(\bfx^{\prime}, \bfy^{\prime}) \in \{0,1\}^{L} \times \{0,1\}^{L}$ as follows.
    \begin{enumerate}
        \item \textbf{Block decomposition.} The symbols of $\bfx$ and $\bfy$ are partitioned sequentially into $M=\frac{2L-18}{3}$ blocks, each consisting of two consecutive symbols from one strand and one symbol from the other, as described in the beginning of this section.
        Let the result be $(B_1, B_2, \ldots, B_{M})$.

        \item \textbf{Permutation of block types.} Apply a permutation $\sigma$ to each block, chosen so that the number of homogeneous blocks is at most $\lfloor M/4 \rfloor$.
        Then, denote it as $(\sigma(B_1), \sigma(B_2), \ldots, \sigma(B_{M}))$.

        \item \textbf{Block flipping.} Optionally flip all even-indexed blocks to ensure that the number of unfriendly boundaries is at most $\lfloor(M-1)/2\rfloor$.

        \item \textbf{Local interleaving.} Each block is interleaved using the greedy rule $g(\cdot)$, and the resulting sequences are concatenated to form a binary sequence
        $$
        \bfz_{\text{payload}} = (b_1, b_2, \ldots, b_{2L-18}).
        $$
        Each symbol in $\bfz_{\text{payload}}$ is assigned to either $\bfx$ or $\bfy$ according to its origin in the block, yielding sequences
        $$
        (x'_1,\ldots,x'_{L-9}) \quad \text{and} \quad (y'_1,\ldots,y'_{L-9}),
        $$
        which preserve the order of $\bfx$ and $\bfy$.

        \item \textbf{Redundancy bits.} Encode $\sigma$ into its $16$-bit representation and set the $\phi$ bit to $1$ if $\phi$ was used, or $0$ otherwise. This results in $17$ redundancy bits, which are encoded to $(r_1, r_2, \ldots, r_{18})$ using the encoder from \cite{Lenz2020}, assuming that the first bit the machine outputs is $1-b_{2L-18}$.
        Append the odd-indexed redundancy bits after $x^{\prime}_{L-9}$ and the even-indexed redundancy bits after $y^{\prime}_{L-9}$.

        \item \textbf{Synthesizing.} The sequences to be synthesized are:
        \begin{align*}
            \bfx^{\prime} &= (x^{\prime}_1, x^{\prime}_2, \ldots, x^{\prime}_{L-9}, r_1, r_3, \ldots, r_{17}), \\
            \bfy^{\prime} &= (y^{\prime}_1, y^{\prime}_2, \ldots, y^{\prime}_{L-9}, r_2, r_4, \ldots, r_{18}).
        \end{align*}
        These sequences are synthesized according to the following interleaving $\bfz$ of $\bfx^{\prime}$ and $\bfy^{\prime}$:
        $$\bfz = (b_1, b_2, \ldots, b_{2L-18}, r_1, r_2, \ldots, r_{18}).$$
    \end{enumerate}

    Given $(\bfx^{\prime}, \bfy^{\prime})$, the decoder extracts the $18$ redundancy bits, decodes the first $17$ bits to recover $\sigma$ and the flipping decision, reconstructs the block structure, reverses the flipping operation, inverts the permutation, and uniquely recovers $(\bfx,\bfy)$.
    Since each block is exactly three bits and the interleaving preserves block boundaries, the block decomposition is uniquely recoverable.

    The bound on the synthesis time of the internals and boundaries of the blocks and the redundancy follow from \Cref{eq:T-payload} and \Cref{eq:T-redundancy}, respectively.
    By construction, the payload part ends with synthesizing $b_{2L-18}$, thus the next bit the machine will output is $1-b_{2L-18}$, and the encoding of the $18$ redundancy bits will assume it when its encoding begins, so no additional boundary cost is incurred.
    Hence, $T_{*}(\bfx^{\prime}, \bfy^{\prime}) \leq T_{\text{pld}} + T_{\text{red}} \leq \frac{8L}{3} + 1$.
\end{IEEEproof}

The theorem guarantees the existence of a feasible schedule, namely the one induced by the construction above. We do not claim that this induced schedule is optimal for the encoded pair $(\bfx^{\prime},\bfy^{\prime})$, and a better schedule may exist.
Moreover, the bound is deterministic and holds for every input pair $(\bfx,\bfy)$, in contrast to the average-case bounds derived in \Cref{sec:expected-synth-time} for uncoded random sequences.
Finally, we provide a detailed example of the encoder described in \Cref{theorem:encoder-binary-synth}.

\begin{example}
    Suppose we are given the following sequences of length $24$:
    \begin{align*}
    \bfx \!&=\! (
    \textcolor{red}{0},\textcolor{red}{0},
    \textcolor{blue}{0},
    \textcolor{red}{0},\textcolor{red}{1},
    \textcolor{blue}{1},
    \textcolor{red}{0},\textcolor{red}{0},
    \textcolor{blue}{0},
    \textcolor{red}{1},\textcolor{red}{0},
    \textcolor{blue}{1},
    \textcolor{red}{1},\textcolor{red}{1},
    \textcolor{blue}{0},
    \textcolor{red}{1},\textcolor{red}{1},
    \textcolor{blue}{1},
    \textcolor{red}{0},\textcolor{red}{0},
    \textcolor{blue}{0},
    \textcolor{red}{1},\textcolor{red}{1},
    \textcolor{blue}{0}
    )
    \\[1ex]
    \bfy \!&=\! (
    \textcolor{red}{0},
    \textcolor{blue}{1},\textcolor{blue}{1},
    \textcolor{red}{1},
    \textcolor{blue}{1},\textcolor{blue}{1},
    \textcolor{red}{0},
    \textcolor{blue}{1},\textcolor{blue}{1},
    \textcolor{red}{1},
    \textcolor{blue}{0},\textcolor{blue}{1},
    \textcolor{red}{1},
    \textcolor{blue}{0},\textcolor{blue}{0},
    \textcolor{red}{1},
    \textcolor{blue}{0},\textcolor{blue}{1},
    \textcolor{red}{1},
    \textcolor{blue}{1},\textcolor{blue}{1},
    \textcolor{red}{0},
    \textcolor{blue}{0},\textcolor{blue}{0}
    ),
    \end{align*}

    We can visualize the split into blocks as follows, where the first row in each block contains the bits from $\bfx$, and the second row contains the bits from $\bfy$:
    \begin{align*}
        B_1 = \begin{bmatrix}
        \textcolor{red}{0} & \textcolor{red}{0} \\
        \textcolor{red}{0} & -
        \end{bmatrix},\,\,
        B_2 = \begin{bmatrix}
        \textcolor{blue}{0} & - \\
        \textcolor{blue}{1} & \textcolor{blue}{1}
        \end{bmatrix},\,\, \ldots,
        B_{16} = \begin{bmatrix}
        \textcolor{blue}{0} & - \\
        \textcolor{blue}{0} & \textcolor{blue}{0}
        \end{bmatrix}.
    \end{align*}
    Note that since $L-9=24$, $L=33$, and $M=\frac{2 \cdot 33 - 18}{3} = 16$.
    Using the formal definition of the blocks, 
    \begin{align*}
        B_1 = \{(\textcolor{red}{0},\textcolor{red}{0}),\,\textcolor{red}{0}\},\;
        B_2 = \{(\textcolor{blue}{1},\textcolor{blue}{1}),\,\textcolor{blue}{0}\},\; \ldots,
        B_{16} = \{(\textcolor{blue}{0},\textcolor{blue}{0}),\,\textcolor{blue}{0}\}.
    \end{align*}
    One can verify that the all-zeros homogeneous block occurs $\frac{4}{16}$ times, and that the all-ones homogeneous block occurs $\frac{3}{16}$ times. That is, the number of homogeneous blocks is $7>4=M/4$. Thus, we find a permutation $\sigma$ that satisfies \Cref{lemma:good-permutation-existance-for-non-homogeneous}. For example, we can select the following permutation:
    \begin{align*}
        \{ (0,0), 0\} &\to \{ (1,0), 1\}, \quad
        \{ (1,1), 1\} \to \{ (0,0), 1\}, \\
        \{ (1,0), 1\} &\to \{ (0,0), 0\}, \quad
        \{ (0,0), 1\} \to \{ (1,1), 1\}, \\
        \{ (0,1),0\} &\to \{ (0,1), 0\}, \quad
        \{ (0,1),1\} \to \{ (0,1), 1\}, \\
        \{ (1,0),0\} &\to \{ (1,0), 0\}, \quad
        \{ (1,1),0\} \to \{ (1,1), 0\}.
    \end{align*}
    It can be verified that now only blocks $7$ and $13$ are homogeneous.
    Furthermore, the concatenated blocks after the permutation have $8$ friendly pairs; thus, we do not need to perform the flipping of even-indexed blocks. 
    Assuming that $(r_1, r_2, \ldots, r_{18})$ are the redundancy bits for the described encoding, 
    \begin{align*}
        \bfx^{\prime}
        =
        (
        &\textcolor{red}{1},\textcolor{red}{0},\textcolor{blue}{1},\textcolor{red}{0},\textcolor{red}{1},\textcolor{blue}{0},
        \textcolor{red}{1},\textcolor{red}{0},\textcolor{blue}{1},\textcolor{red}{0},\textcolor{red}{0},\textcolor{blue}{0},
        \textcolor{red}{0},\textcolor{red}{0},\textcolor{blue}{0},\textcolor{red}{0},\textcolor{red}{0},\textcolor{blue}{0}, \\
        &\textcolor{red}{1},\textcolor{red}{1},\textcolor{blue}{1},\textcolor{red}{1},\textcolor{red}{1},\textcolor{blue}{0},r_1,r_3,\ldots,r_{17}
        )
    \end{align*}
    \begin{align*}
        \bfy^{\prime}
        =
        (
        &\textcolor{red}{1},\textcolor{blue}{0},\textcolor{blue}{0},\textcolor{red}{1},\textcolor{blue}{1},\textcolor{blue}{1},
        \textcolor{red}{1},\textcolor{blue}{0},\textcolor{blue}{0},\textcolor{red}{0},\textcolor{blue}{1},\textcolor{blue}{0},
        \textcolor{red}{1},\textcolor{blue}{0},\textcolor{blue}{1},\textcolor{red}{1},\textcolor{blue}{1},\textcolor{blue}{0}, \\
        &\textcolor{red}{1},\textcolor{blue}{0},\textcolor{blue}{0},\textcolor{red}{0},\textcolor{blue}{0},\textcolor{blue}{1},r_2,r_4,\ldots,r_{18}
        )
    \end{align*}
\end{example}

\section{Optimal Synthesis Order}\label{sec:optimal-order}
%
%

In this section, we address \Cref{problem:finding-optimal-synthesis-order} and characterize the optimal synthesis order for two fixed sequences $\bfx, \bfy \in \Sigma_q^L$.

At any time slot $t$, the system is fully described by the triple $(i,j,r)$, where:
$i$ is the number of symbols already synthesized in $\bfx$,
$j$ is the number of symbols already synthesized in $\bfy$,
and $r \in \Sigma_q$ is the symbol currently emitted by the periodic synthesis sequence, i.e., $r = t \bmod q$.

Define the DP array of size $(L+1) \times (L+1) \times q$, where indexing is zero-based, and let $\text{DP}(i,j,r)$ be the minimum number of additional time slots required to complete both sequences starting from state $(i,j,r)$.
The goal is to compute $\text{DP}(L,L,r)=0$.

Let $r^{\prime} = (r+1) \pmod q$ represent the next symbol to be synthesized.
Then, for all $0 \leq i,j \leq L$ and $r \in \Sigma_{q}$, the recurrence is defined as follows:\par\noindent
\begin{footnotesize}\begin{align*}
    \text{DP}(i,j,r) \!=\! 
    \begin{cases} 
        \begin{aligned}
            \min\{ &1 + \text{DP}(i+1,j,r'), \\
            &1 + \text{DP}(i,j+1,r')\} 
        \end{aligned} & \text{if } x_{i+1} = y_{j+1} = r, \\
        1 + \text{DP}(i+1,j,r') & \text{if } i \!<\! L, x_{i+1} \!=\! r  \!\neq\! y_{j+1}, \\
        1 + \text{DP}(i,j+1,r') & \text{if } j \!<\! L, y_{j+1} \!=\! r  \!\neq\! x_{i+1}, \\
        1 + \text{DP}(i,j,r') & \text{if } x_{i+1} \neq r, y_{j+1} \neq r, \\
        0 & \text{if } i = L, j=L.
    \end{cases}
\end{align*}\end{footnotesize}
Undefined symbols (when $i \geq L$ or $j \geq L$) are treated as unequal to any $r$.
The cases correspond directly to the possible actions in the current time slot:
\begin{enumerate}
    \item Both sequences can synthesize ($x_{i+1}=y_{j+1}=r$); one chooses the better option.
    \item Only $\bfx$ can synthesize.
    \item Only $\bfy$ can synthesize.
    \item Neither can synthesize; the algorithm must wait for the next cycle.
\end{enumerate}
The terminal condition $\text{DP}(L,L,r)=0$ reflects that both sequences are already complete.
Note that only (4) results in an IDLE action, as derived in \Cref{lemma:greedy-is-opt}.
The correctness of this recurrence is established in \Cref{theorem:DP-rec-correctness}.
\begin{theorem}
\label{theorem:DP-rec-correctness}
    $T_{*}(\bfx,\bfy) = \text{DP}(0,0,0)$.
\end{theorem}
\begin{IEEEproof}
First, note that in all cases we assume that ``IDLE'' action is chosen only when there is no other action possible, and it is optimal, as we showed in \Cref{lemma:greedy-is-opt}.
We prove by induction on $\ell = (L-i)+(L-j)$, i.e., the total number of symbols remaining to synthesize in both sequences together.
The base case, $\ell=0$ ($i=j=L$), is trivial and is defined to be $\text{DP}(L,L,r)=0$ for all $r$.
For the inductive step, we assume it holds for all $\ell^{\prime}\!<\!\ell$, for any $0 \leq r \leq q\!-\!1$, and we split it into cases.
Let $\sigma \!=\! r$ be the currently available symbol.

\begin{itemize}
    \item Case 1: $i < L$, $j < L$, $x_{i+1} = \sigma$, and $y_{j+1} = \sigma$.
    Then, the minimum number of cycles is obtained either by synthesizing $x_{i+1}$ or $y_{j+1}$.
    If we choose $x_{i+1}$, $\text{DP}(i+1,j,r^{\prime})$ is correct according to the induction hypothesis; thus, the minimum number of cycles will be $1+\text{DP}(i+1,j,r^{\prime})$.
    Similarly, for choosing $y_{j+1}$, we find that the minimal number of cycles is $1+\text{DP}(i,j+1,r^{\prime})$.
    Thus, we conclude that $\text{DP}(i,j,r) = 1 + \min\{\text{DP}(i+1,j,r^{\prime}), \text{DP}(i,j+1,r^{\prime})\}$.

    \item Case 2: $i<L$, $x_{i+1} = \sigma$, and $y_{j+1} \neq \sigma$.
    Then, the only possible advance is to synthesize $x_{i+1}$. By the induction hypothesis, the remaining cost is $\text{DP}(i+1,j,r^{\prime})$, so $\text{DP}(i,j,r)=1+\text{DP}(i+1,j,r^{\prime})$.

    \item Case 3: $j<L$, $y_{j+1} = \sigma$, and $x_{i+1} \neq \sigma$. This case is symmetric to case 2.

    \item Case 4: $x_{i+1} \neq \sigma$ and $y_{j+1} \neq \sigma$ (and at least $i$ or $j$ are smaller than $L$), then we cannot synthesize either sequences' next symbol; thus, $\text{DP}(i,j,r) = 1 + \text{DP}(i,j,r^{\prime})$.
    There is an important consideration here. We know that there exists a symbol $\sigma$ for which $x_{i+1}$ or $y_{j+1}$ is equal to $\sigma$. 
    Thus, this recursion of $\text{DP}(i,j,r) = 1 + \text{DP}(i,j,r^{\prime})$ converges by one of the previous cases, since each skip operation preserves optimality (waiting does not create better opportunities than those that will naturally arise). 
\end{itemize}

The minimum value computed through this chain of skips, followed by eventual progress, represents the true optimal synthesis time.
To conclude, $\text{DP}(0,0,0)$ correctly computes the minimum time to reach $\text{DP}(L,L,\cdot)$, which corresponds to having synthesized all the symbols from each sequence.
\end{IEEEproof}

We derived \Cref{algorithm:min-synth-seq}, a dynamic programming algorithm  that solves it with $O(L^2 q)$ time and space complexity.

\begin{algorithm}
    \caption{Algorithm for finding the minimum synthesis time of $\bfx$ and $\bfy$ of length $L$.}
    \label{algorithm:min-synth-seq}
    \begin{algorithmic}[1]
        \renewcommand{\algorithmicrequire}{\textbf{Input:}}
        \REQUIRE Alphabet $\Sigma_q$, $\bfx$, and $\bfy$.
        \renewcommand{\algorithmicensure}{\textbf{Output:}}
        \ENSURE The minimum synthesis time of $\bfx$ and $\bfy$.
        \FOR {( $r = 0$ to $q-1$ )}
        	\STATE $\text{DP}[L][L][r] \leftarrow 0$.
        \ENDFOR
        \FOR {($i = L$ down to $0$)}
        	\FOR {($j = L$ down to $0$)}
                \IF {($i = L$ and $j = L$)}
        			\STATE continue
                \ENDIF
                
        		\STATE startR $\leftarrow$ findFirstProgressSymbol($\bfx$, $\bfy$, $i$, $j$).
        		
        		\FOR {($k = 0$ to $q-1$)}
                    \STATE $r \leftarrow (\text{startR} - k)\,\bmod\,q$.
                    \STATE $r^{\prime} \leftarrow (r+1)\,\bmod\,q$.
        			\STATE canX $\leftarrow (i < L)$ and $(\bfx[i] = r)$.
                    \STATE canY $\leftarrow (j < L)$ and $(\bfy[j] = r)$.
        			\IF {(canX and canY)}
        		      \STATE val $\leftarrow 1 + \min\{\text{DP}[i+1][j][r^{\prime}], \text{DP}[i][j+1][r^{\prime}]\}$.
                    \ELSIF{(canX)}
                        \STATE val $\leftarrow 1 + \text{DP}[i+1][j][r^{\prime}]$.
                    \ELSIF{(canY)}
                        \STATE val $\leftarrow 1 + \text{DP}[i][j+1][r^{\prime}]$.
                    \ELSE
                        \STATE val $\leftarrow 1 + \text{DP}[i][j][r^{\prime}]$.
                    \ENDIF
        			\STATE $\text{DP}[i][j][r] \leftarrow$ val.
                \ENDFOR
            \ENDFOR
        \ENDFOR
        \RETURN $\text{DP}[0][0][0]$.
    \end{algorithmic} 
\end{algorithm}

Note that \Cref{algorithm:min-synth-seq} uses the following auxiliary function, \texttt{findFirstProgressSymbol($\bfx$, $\bfy$, $i$, $j$)}, which returns the first symbol $r \in \Sigma_q$ such that either $x_i = r$ or $y_j = r$ (i.e., the first symbol that allows progression from state $(i,j)$).
That is, the first time slot (starting at a time slot where $r=0$) in which a non-IDLE action can occur.
The loop over $k$ iterates backward over the cyclic alphabet, starting from the first symbol that enables progress, so that all $q$ possible symbols $r$ are evaluated in the correct cyclic order for the current state.


\section*{Acknowledgments}
The authors thank Ohad Elishco for carefully reading an earlier version of the manuscript, identifying issues in several proofs, and suggesting \Cref{lemma:limit-of-gamma-exists}. The authors also thank Mark Somoza for helpful discussions on the synthesis model of photolithographic synthesis. 
The research was funded by the European Union (ERC, DNAStorage, 101045114, and EIC, DiDAX 101115134). Views and opinions expressed are, however, those of the authors only and do not necessarily reflect those of the European Union or the European Research Council Executive Agency. Neither the European Union nor the granting authority can be held responsible for them.
This work was also supported in part by NSF Grant CCF2212437.

\appendix\label{appendix}

\greedisopt*
\begin{IEEEproof}
For simplicity, we will prove for $k=2$. The generalization for larger $k$ can be done with similar arguments.
Consider the $1 \times 2$ synthesis model with a fixed alternating sequence $(r_t)_{t \geq 0}$ over $\Sigma_q$, and two strands $\bfx$ and $\bfy$, synthesized in order.
We allow policies to be as general as possible; they may idle even when one or both strands could advance.
We will show that there exists a policy achieving $T_{*}$ that never idle-waits when progress is possible.
At time slot $t$, let $i_t$ be the number of symbols of $\bfx$ synthesized up to (but not including) time slot $t$, and $j_t$ be the same for $\bfy$.

We say that an action $a_t$ of a schedule $\bfa=(a_1,a_2,\ldots,a_T)$ is a \emph{bad idle} if at least one strand can advance, i.e.,
$$(i_t < L \text{ and } x_{i_t + 1}=r_t) \text{ or } (j_t < L \text{ and } y_{j_t + 1}=r_t),$$
but the action chosen is $a_t=\text{IDLE}$.
Define, for a schedule $\bfa$,
\begin{align*}
	E(\bfa) \coloneqq
	\begin{cases}
		\min \{t \colon a_t \text{ is a bad idle} \}  & \text{if any bad idle exists}, \\
		\infty & \text{if no bad idles exist}.
	\end{cases}
\end{align*}
Denote $T(\bfa)$ as the completion time of a specific schedule and consider the set of all optimal schedules:
$$\cO \coloneqq \{ \bfa \colon T(\bfa)=T_{*} \}.$$
If there is some $\bfa \in \cO$ with $E(\bfa)=\infty$, we are done.
So suppose, for contradiction, that every optimal schedule has at least one bad idle, i.e., $E(\bfa)<\infty$ for all $\bfa \in \cO$.
Among all $\bfa \in \cO$, choose one with the latest first bad idle:
$$\bfa^{*} \in \argmax_{\bfa \in \cO} E(\bfa).$$
Let $t_0 \coloneqq E(\bfa^{*})$ be its earliest bad idle index; by assumption, $t_0 < \infty$.
By the definition of a bad idle at $t_0$, we have $a^{*}_{t_0} = \text{IDLE}$, and at least one of the next symbols matches the current synthesis symbol:
$$(i_{t_0} < L \text{ and } x_{i_{t_0}+1}=r_{t_0}) \text{ or } (j_{t_0} < L \text{ and } y_{j_{t_0}+1} = r_{t_0}).$$
Also, since $t_0$ is the earliest bad idle of $\bfa^{*}$, there are no bad idles at times $t < t_0$.
We distinguish two cases at time $t_0$.
Let the state at slot $t_0$ be $(i_0,j_0,r_0)$ with $i_0 \coloneqq i_{t_0}$, $j_0 \coloneqq j_{t_0}$, and $r_0 \coloneqq r_{t_0}$.
Case A (single feasible strand) is the case where exactly one strand can advance at $t_0$.
W.l.o.g., suppose it is $\bfx$, and either $j_0=L$ or $y_{j_0+1} \neq r_0$.
Case B (tie) is where both strands can advance at $t_0$.
We now treat both cases in a unified way by selecting one particular strand whose missed advance at $t_0$ we will move forward to $u>t_0$.

In Case A, the choice is forced: we select $\bfx$. In Case B, we select the strand that is first actually advanced after $t_0$ by $\bfa^{*}$. Again, w.l.o.g., suppose it is $\bfx$.
We obtain an index $u>t_0$ with the following properties:
at time $t_0$, $\bfx$ could have advanced (its next symbol equals $r_0$);
$u$ is the first time after $t_0$ that $\bfa^{*}$ actually advances $\bfx$;
and on the interval $(t_0,u)$, $\bfx$ is never advanced in $\bfa^{*}$.
Now define a new schedule $\Tilde{\bfa}$ by modifying $\bfa^{*}$ only on the interval $[t_0,u]$ as follows.
For all $t<t_0$, set $\Tilde{\bfa}_t \coloneqq a^{*}_t$.
At time $t_0$, set $\Tilde{\bfa}_{t_0}$ to ``advance $\bfx$'' instead of IDLE; this is allowed because, by construction, $\bfx$ can advance at $(i_0,j_0,r_0)$.
For all $t_0 < t < u$, set $\Tilde{\bfa}_t \coloneqq a^{*}_t$; note that on this interval $\bfa^{*}$ never advanced $\bfx$, so we are only copying IDLE or advances of $\bfy$.
At time $u$, set $\Tilde{\bfa}_u \coloneqq \text{IDLE}$ instead of an advance of $\bfx$.
For all $t > u$, set $\Tilde{\bfa}_t \coloneqq a^{*}_t$.
Intuitively, we have swapped an idle at $t_0$ with the first later advance of $\bfx$; i.e., we advance $\bfx$ earlier and idle later.

We must verify that $\Tilde{\bfa}$ respects all constraints.
At $t_0$: by construction $\bfx$ could advance, so the new action is feasible.
For $t_0 < t < u$: in $\bfa^{*}$, $\bfx$ was never advanced during this interval (by choice of $u$); in $\Tilde{\bfa}$, $\bfx$ has one additional advance at $t_0$, but since we never attempt to advance $\bfx$ in $(t_0,u)$, the feasibility conditions for all actions on $(t_0,u)$ are unaffected.
At time $u$: we replaced an advance of $\bfx$ with IDLE, which is always allowed.
For $t>u$: the pattern of actions is identical to $\bfa^{*}$; since each strand is synthesized in order and we used exactly the same number of advances for each strand, future actions remain feasible.
Thus, $\Tilde{\bfa}$ is a valid schedule.

Since we performed exactly the same number of idles, advances of $\bfx$, and advances of $\bfy$ as in $\bfa^{*}$---just redistributed within $[t_0,u]$---the time at which both strands are fully synthesized is unchanged, so $\Tilde{\bfa} \in \cO$.

Finally, $\Tilde{\bfa}$ agrees with $\bfa^{*}$ before $t_0$ and does not idle at $t_0$, so any bad idle in $\Tilde{\bfa}$ occurs at some time $>t_0$, giving $E(\Tilde{\bfa}) > t_0 = E(\bfa^{*})$.
This contradicts the maximality of $\bfa^{*}$, so there must exist some $\bfa \in \cO$ with $E(\bfa)=\infty$.
\end{IEEEproof}


\claimETEX*
\begin{IEEEproof}
    Index the slots of one full rotation by $t=0,1,\ldots,T-1$.
    Thus, $a_0=0$ and $a_T=0$, where $a_T$ is the start of the next full rotation.
    Let $\Delta a_t \coloneqq a_{t+1}-a_t$.
    Then,
    \begin{equation}
        \sum_{t=0}^{T-1}\Delta a_t = a_T-a_0=0. \label{eq:sum-delta-at-is-zero}
    \end{equation}
    Let $\mathcal{F}_t$ denote the information available up to the beginning of slot $t$. Define 
    $$M_n \coloneqq \sum_{t=0}^{n-1} \left( \Delta a_t-\mathbb{E}[\Delta a_t\mid \mathcal{F}_t]
    \right).$$
    Then, $(M_n)_{n\ge0}$ is a martingale, since $M_{n+1}=M_n+(\Delta a_n - \mathbb{E}[\Delta a_n \mid \mathcal{F}_n])$, and by taking expectation given $\mathcal{F}_n$:
    \begin{align*}
    	\mathbb{E}[M_{n+1}\mid \mathcal{F}_n] &= \mathbb{E}[M_{n}\mid \mathcal{F}_n] + \mathbb{E}[\Delta a_n - \mathbb{E}[\Delta a_n \mid \mathcal{F}_n] \mid \mathcal{F}_n] \\
        &= M_{n} + \mathbb{E}[\Delta a_n \mid \mathcal{F}_n] - \mathbb{E}[\mathbb{E}[\Delta a_n \mid \mathcal{F}_n] \mid \mathcal{F}_n] \\
        &= M_{n} + \mathbb{E}[\Delta a_n \mid \mathcal{F}_n] - \mathbb{E}[\Delta a_n \mid \mathcal{F}_n] = M_n.
    \end{align*}
    Moreover, $T$ is a stopping time, and by \Cref{lemma:T-has-exponential-tail} it has finite expectation. 
    Since $\Delta a_t$ is bounded, $\mathbb{E}[M_T]=\mathbb{E}[M_0]=0$ by the optional stopping theorem~\cite[Theorem 3.6]{Bhattacharya2007-ca}.
    Therefore,
    $$\mathbb{E}\left[\sum_{t=0}^{T-1}\Delta a_t\right] = \mathbb{E}\left[ \sum_{t=0}^{T-1}
    \mathbb{E}[\Delta a_t\mid \mathcal{F}_t] \right].$$
    Since the left-hand side is zero by \Cref{eq:sum-delta-at-is-zero}, we get
    \begin{equation}
        0= \mathbb{E}\left[\sum_{t=0}^{T-1} \mathbb{E}[\Delta a_t\mid \mathcal{F}_t]\right]. \label{eq:expected-sum-delta-at-is-zero}
    \end{equation}
    Next, we compute the conditional expectation. 
    If $a_t=0$, then $\bfx$ advances and the next offset is uniform on $\{0,1,\ldots,q-1\}$. Hence,
    $$\mathbb{E}[\Delta a_t \mid \mathcal{F}_t] = \frac{q-1}{2}.$$
    If $a_t>0$, then $a_{t+1}=a_t-1$, so
    $$\mathbb{E}[\Delta a_t\mid \mathcal{F}_t]=-1.$$
    Therefore,
    $$\mathbb{E}[\Delta a_t\mid \mathcal{F}_t] = \frac{q-1}{2} \bfone_{\{a_t=0\}} - \bfone_{\{a_t>0\}}.$$
    Substituting into \Cref{eq:expected-sum-delta-at-is-zero} gives
    $$0=\mathbb{E}\left[ \sum_{t=0}^{T-1} \left( \frac{q-1}{2}\bfone_{\{a_t=0\}} - \bfone_{\{a_t>0\}} \right) \right].$$
    Since $\bfone_{\{a_t>0\}}=1-\bfone_{\{a_t=0\}}$, we obtain
    $$0 = \mathbb{E}\left[ \sum_{t=0}^{T-1} \left( \frac{q+1}{2} \bfone_{\{a_t=0\}}-1 \right) \right].$$
    Thus,
    $$0=\frac{q+1}{2} \mathbb{E}\left[ \sum_{t=0}^{T-1}\mathbf 1_{\{a_t=0\}} \right] - \mathbb{E}[T].$$
    By definition, the indicator counts exactly $V_X$. Therefore,
    $$0=\frac{q+1}{2}\mathbb{E}[V_X]-\mathbb{E}[T],$$
    and hence
    $$\mathbb{E}[T]=\frac{q+1}{2}\mathbb{E}[V_X].$$
\end{IEEEproof}

\claimEXEY*
\begin{IEEEproof}
As in \Cref{claim:E[T]-and-E[X]}, index the slots of one full rotation by $t=0,1,\ldots,T-1$.
Let $d_{t} \coloneqq a_{t} - b_{t}$ denote the drift, $\Delta d_t:=d_{t+1}-d_t$.
Since the full rotation starts and ends at $(0,0)$, we have $d_0=d_T=0$,
and therefore
$$\sum_{t=0}^{T-1}\Delta d_t=0.$$
As in \Cref{claim:E[T]-and-E[X]}, let $\mathcal{F}_t$ denote the information available up to the beginning of slot $t$, and define
$$ M_n \coloneqq \sum_{t=0}^{n-1} \left( \Delta d_t-\mathbb{E}[\Delta d_t\mid\mathcal{F}_t] \right). $$
Then $(M_n)_{n\geq0}$ is a martingale, by a similar argument to the one in \Cref{claim:E[T]-and-E[X]}. 
Since $T$ is a stopping time with finite expectation by \Cref{lemma:T-has-exponential-tail}, and $\Delta d_t$ is bounded, the optional stopping theorem~\cite[Theorem 3.6]{Bhattacharya2007-ca} gives $\mathbb E[M_T]=0$.
Hence,
\begin{equation}
    0 = \mathbb{E}\left[ \sum_{t=0}^{T-1} \mathbb{E}[\Delta d_t\mid\mathcal{F}_t] \right]. \label{eq:expected-drift-sum-is-zero}
\end{equation}
We now compute the conditional expectation according to the current state.
If $a_t=b_t=0$, then a tie occurs and $\bfx$ advances, so $d_{t+1}=a_{t+1}-b_{t+1}$, where $a_{t+1}$ is uniform over $\{0,1,\ldots,q-1\}$ and $b_{t+1}=q-1$.
Therefore,
$$\mathbb{E}[\Delta d_t\mid\mathcal{F}_t] = -\frac{q-1}{2}.$$
If $a_t=0$ and $b_t>0$, then only $\bfx$ advances.
Thus, $d_{t+1}=a_{t+1}-(b_t-1)$, where $a_{t+1}$ is uniform over $\{0,1,\ldots,q-1\}$, and hence
$$\mathbb{E}[\Delta d_t\mid\mathcal{F}_t] = \frac{q+1}{2}.$$
Similarly, if $a_t>0$ and $b_t=0$, then only $\bfy$ advances, yielding
$$\mathbb{E}[\Delta d_t\mid\mathcal{F}_t] = -\frac{q+1}{2}.$$
Finally, if $a_t>0$ and $b_t>0$, then neither strand resets, and
$d_{t+1}=(a_t-1)-(b_t-1)=d_t$, so
$$\mathbb{E}[\Delta d_t\mid\mathcal{F}_t]=0.$$
Therefore,
\begin{align*}
    \mathbb{E}[\Delta d_t\mid\mathcal{F}_t] &= \frac{q+1}{2}\bfone_{\{a_t=0,b_t>0\}} \\
    &- \frac{q+1}{2}\bfone_{\{a_t>0,b_t=0\}} - \frac{q-1}{2}\bfone_{\{a_t=b_t=0\}}.
\end{align*}
Substituting into \Cref{eq:expected-drift-sum-is-zero}:
$$0= \frac{q+1}{2} \mathbb{E}[V_X-1] - \frac{q+1}{2}\mathbb{E}[V_Y] - \frac{q-1}{2},$$
where the term $V_X-1$ excludes the initial tie.
Rearranging,
\begin{align*}
    0 &= \frac{q+1}{2} \mathbb{E}[V_X] -\frac{q+1}{2} - \frac{q+1}{2}\mathbb{E}[V_Y] - \frac{q-1}{2} \\
    0 &=\frac{q+1}{2} \left( \mathbb{E}[V_X] - \mathbb{E}[V_Y] \right) - q.
\end{align*}
Thus,
$$\mathbb{E}[V_X]-\mathbb{E}[V_Y] = \frac{2q}{q+1}.$$
\end{IEEEproof}

\claimEXEYFinal*
\begin{IEEEproof}
    We show only $\mathbb{E}[V_X] = \frac{q(q+3)}{2(q+1)}$ since the rest are direct conclusions drawn from applying \Cref{claim:E[T]-and-E[X]} and \Cref{claim:E[X]-E[Y]}.
    Let $g(a,b)$ for $a,b\in \{0,1,\ldots, q-1\}$ be the expected number of $\bfx$-visits until the next arrival to state $(0,0)$, i.e., the next ``tie,'' which is the end of the full rotation.
    
    Let $A_b \coloneqq g(0,b)$ and $B_a \coloneqq g(a,0)$ for all $1 \leq a, b \leq q-1$, where $A_b$ denotes the expected number of $\bfx$-visits, starting from a state of the form $(0,b)$, and $B_a$ denotes the expected $\bfx$-visits from a current $\bfy$-advance state.
    The value $g(0,0)$ refers to the beginning of a new full rotation. When $(0,0)$ is reached from another state, the current full rotation terminates. Thus, in the recurrences below, we use the convention $A_0=B_0=0$ and wish to compute $g(0,0) = \mathbb{E}[V_X]$.
    Note that if $a,b \geq 1$, then the choice of the next state is deterministic, and it is $(a-1,b-1)$.
    Hence, the following holds:
    $$
        g(a,b) = 
        \begin{cases}
            \mathbb{E}[V_X], & \text{if } a=b=0,\\
            A_b, & \text{if } a=0<b,\\
            B_a, & \text{if } b=0<a,\\
            A_{b-a}, & \text{if } 0<a<b,\\
            B_{a-b}, & \text{if } 0<b<a,\\
            0, & \text{otherwise } (0<a=b).
        \end{cases}
    $$
    At the beginning of a full rotation, the initial tie is resolved in favor of $\bfx$.
    Hence, after one $\bfx$-visit, the next state is $(u,q-1)$, where $u\sim\mathrm{Unif}\{0,\ldots,q-1\}$.
    Thus, it holds that,
    \begin{align*}
        \mathbb{E}[V_X] &= g(0,0) = 1 + \frac{1}{q} \sum_{u=0}^{q-1} g(u,q-1) \nonumber \\
        &= 1 + \frac{1}{q} \left( \sum_{u=0}^{q-2} g(u,q-1) + g(q-1,q-1) \right) \nonumber.
    \end{align*}
    Since $g(q-1,q-1)=0$ and $g(u,q-1) = A_{q-1-u}$, we arrive at a result through a change of index and reordering of the summation, as follows:
    \begin{equation}
        \mathbb{E}[V_X] = 1+\frac{1}{q}\sum_{u=0}^{q-2} A_{q-1-u} = 1+\frac{1}{q}\sum_{r=1}^{q-1} A_r.\label{eq:Ex-as-sum-of-Ar}
    \end{equation}
    Similarly,
    \begin{align*}
        A_b &= 1 + \frac{1}{q} \left( \sum_{u=0}^{b-1} A_{b-1-u} + \sum_{u=b}^{q-1} B_{u-(b-1)} \right), \\
        B_a &= \frac{1}{q} \left( \sum_{v=0}^{a-1} B_{a-1-v} + \sum_{v=a}^{q-1} A_{v-(a-1)} \right).
    \end{align*}
    Note that in these expressions, $A_0 = B_0 = 0$ is used by convention. For $q=2$, it is: 
    $$A_1=1+\frac{1}{2}B_1,\qquad B_1=\frac{1}{2}A_1,$$
    and hence $A_1=4/3$ and $B_1=2/3$.
    Thus, 
    $$\mathbb{E}[V_X]=1+\frac{1}{2}A_1=\frac{5}{3} = \frac{q(q+3)}{2(q+1)}.$$
    It remains to consider $q \geq 3$. By \Cref{lemma:recurrence-solution}, the solution to these recurrences is:
    $$A_b=\frac{b}{q+1}+\frac{q}{2}, \qquad B_a=-\frac{a}{q+1}+\frac{q}{2}.$$
    Substituting this into \Cref{eq:Ex-as-sum-of-Ar} and using the sum of an arithmetic series,
    \begin{align*}
        \mathbb{E}[V_X] &= 1 + \frac{1}{q} \left( \frac{q(q-1)}{2(q+1)} + \frac{q}{2}(q-1) \right) \\
        &= 1 + \frac{q-1}{2(q+1)} + \frac{q-1}{2} \\
        &= \frac{2q + 2 + q - 1 + q^2 - 1}{2(q+1)} = \frac{q(q+3)}{2(q+1)}.
    \end{align*}
\end{IEEEproof}

We next use the preceding full-rotation analysis to evaluate the finite-length performance of the $\bfx$-first policy.
For the $i$-th full rotation, let $(T_i, V_{X,i}, V_{Y,i})$ denote its length, the number of
$\bfx$-visits, and the number of $\bfy$-visits, respectively, regardless of the tie-breaking policy.
Define
$$X_{\mathrm{tot}}(n) \coloneqq \sum_{i=1}^n V_{X,i},\qquad
Y_{\mathrm{tot}}(n) \coloneqq \sum_{i=1}^n V_{Y,i},$$
with $X_{\mathrm{tot}}(0) = Y_{\mathrm{tot}}(0) = 0$.

\begin{lemma}\label{lemma:recurrence-solution}
Let $A_b$ and $B_a$, where $1 \leq a,b \leq q-1$, satisfy the recurrences
\begin{align}
    A_b &= 1 + \frac{1}{q} \left( \sum_{u=0}^{b-1} A_{b-1-u} + \sum_{u=b}^{q-1} B_{u-(b-1)} \right), \label{eq:A_b} \\
    B_a &= \frac{1}{q} \left( \sum_{v=0}^{a-1} B_{a-1-v} + \sum_{v=a}^{q-1} A_{v-(a-1)} \right). \label{eq:B_a}
\end{align}
with $A_0=B_0=0$.
Then, for every $1 \leq b \leq q-1$,
$$A_b=\frac{b}{q+1}+\frac{q}{2},$$
and for every $1 \leq a \leq q-1$,
$$B_a=-\frac{a}{q+1}+\frac{q}{2}.$$
\end{lemma}
\begin{IEEEproof}
    For $b \geq 2$, multiply \Cref{eq:A_b} by $q$ to obtain the following:
    \begin{align*}
        q \cdot A_b &= q + A_{b-1} + \sum_{u=1}^{b-1} A_{b-1-u} + \sum_{u=b}^{q-1} B_{u-(b-1)}, \\
        q \cdot A_{b-1} &= q + A_{b-2} + \sum_{u=1}^{b-2} A_{b-2-u} + \sum_{u=b-1}^{q-1} B_{u-(b-2)}.
    \end{align*}
    Subtract, telescope, and do the same for \Cref{eq:B_a} where $a \geq 2$ to get
    \begin{align}
        q(A_{b} - A_{b-1}) = A_{b-1} - B_{q-b+1}, \label{eq:A_(b)-A_(b-1)} \\
        q(B_{a} - B_{a-1}) = B_{a-1} - A_{q-a+1}. \label{eq:B_(a)-B_(a-1)}
    \end{align}
    Extracting $B_{q-b+1}$ and $B_{q-b}$ from \Cref{eq:A_(b)-A_(b-1)}, 
    \begin{align}
       B_{q-b+1} = (1+q)A_{b-1} - qA_{b}, \label{eq:B_(q-b+1)} \\
        B_{q-b} = (1+q)A_{b} - qA_{b+1}. \label{eq:B_(q-b)}
    \end{align}
    Using \Cref{eq:B_(a)-B_(a-1)} for $a=q-b+1$ yields,
    \begin{align}
        q(B_{q-b+1} - B_{q-b}) = B_{q-b} - A_{b}. \label{eq:B_(q-b+1) - B_(q-b)}
    \end{align}
    Substituting \Cref{eq:B_(q-b+1)} and \Cref{eq:B_(q-b)} in the LHS of \Cref{eq:B_(q-b+1) - B_(q-b)},
    $$
        q(B_{q-b+1} - B_{q-b}) = q ( (1+q) A_{b-1} - (2q+1) A_b + q A_{b+1} ).
    $$
    And substituting \Cref{eq:B_(q-b)} in the RHS of \Cref{eq:B_(q-b+1) - B_(q-b)},
    $$
        B_{q-b} - A_{b} = q ( A_b - A_{b+1} ).
    $$
    Equate RHS and LHS, and divide by $q$; then the following holds:
    \begin{gather}
        (1+q) A_{b-1} - (2q+1) A_b + q A_{b+1} = A_b - A_{b+1} \nonumber \\
        (q+1) A_{b-1} - 2(q+1) A_b + (q+1) A_{b+1} = 0 \nonumber
    \end{gather}
    For the interior values of $b$, this gives
    \begin{equation}
        A_{b+1} = 2A_{b} - A_{b-1}. \label{eq:A_b-recurrence}
    \end{equation}
    Since \Cref{eq:A_b-recurrence} is a linear recurrence of order 2 with constant coefficients, and its characteristic polynomial has one root of multiplicity $2$, which is $\lambda=1$, hence $A_b$ is affine in $b$, say,
    \begin{equation}
        A_{b} = \alpha b + \beta. \label{eq:A_b-general-sol}
    \end{equation}
    Next, take \Cref{eq:B_(a)-B_(a-1)} and substitute \Cref{eq:A_b-general-sol} to get,
    $$q B_a - (q+1) B_{a-1} = -\alpha (q-a+1) - \beta.$$
    Guess that $B_a = ua + v$, and then, 
    $$qu + u -ua -v = -\alpha (q-a+1) - \beta.$$
    By equating the coefficient of $a$, we find $u = -\alpha$ and use it to equate the constants $v = \beta$.
    Thus,
    \begin{equation}
        B_{a} = -\alpha a + \beta. \label{eq:B_a-general-sol}
    \end{equation}
    To find $\alpha$ and $\beta$, we compute two initial conditions directly, which are $A_1$ and $B_1$.
    From \Cref{eq:A_b} and \Cref{eq:B_a}, 
    $$
        A_1 = 1+ \frac{1}{q} \sum_{a=1}^{q-1} B_{a}, \quad B_1 = \frac{1}{q} \sum_{b=1}^{q-1} A_{b}.
    $$
    Applying \Cref{eq:A_b-general-sol} and \Cref{eq:B_a-general-sol}, and summing an arithmetic series,
    \begin{align*}
        A_1 &= 1+ \frac{1}{q} \left( -\alpha\frac{q(q-1)}{2} + \beta (q-1) \right), \\
        B_1 &= \frac{1}{q} \left( \alpha\frac{q(q-1)}{2} + \beta (q-1) \right).
    \end{align*}
    Also, \Cref{eq:A_b-general-sol} and \Cref{eq:B_a-general-sol} give $A_1 = \alpha + \beta$ and $B_1 = -\alpha + \beta$.
    Therefore, a system of two linear equations is derived:
    \begin{align*}
        \alpha + \beta &= 1 -\alpha \frac{q-1}{2} + \beta \frac{q-1}{q}, \\
        -\alpha + \beta &= \alpha \frac{q-1}{2} + \beta \frac{q-1}{q}.
    \end{align*}
    Solving it yields $\alpha = \frac{1}{q+1}$ and $\beta = \frac{q}{2}$.
    Applying it to \Cref{eq:A_b-general-sol}, it holds that,
    \begin{align*}
        A_b = \frac{b}{q+1} + \frac{q}{2}.
    \end{align*}
    Similarly, substituting $\alpha=\frac{1}{q+1}$ and $\beta=\frac{q}{2}$ into \Cref{eq:B_a-general-sol} gives
    \begin{align*}
        B_a=-\frac{a}{q+1}+\frac q2.
    \end{align*}
    This completes the proof.
\end{IEEEproof}

\propxfirst*
\begin{IEEEproof}
Since each full rotation starts at $(0,0)$, the tie is always resolved in favor of $\bfx$, and the future unread symbols are independent of the past, the triples $(V_{X,i},V_{Y,i},T_i)$, $i \geq 1$, are i.i.d. copies of $(V_X,V_Y,T)$.
Define the stopping time $N_X(L)\coloneqq\min\{n:X_{\mathrm{tot}}(n)\ge L\}$.
That is, $N_X(L)$ is the first full rotation by which we have accumulated at least $L$ $\bfx$-visits.
Let $\kappa$ be the number of symbols of $\bfy$ synthesized by the time $\bfx$ is completed. 
Since $\bfx$ completes during full rotation $N_X(L)$,
\begin{equation}
    Y_{\mathrm{tot}}(N_X(L)-1) \leq \kappa \leq Y_{\mathrm{tot}}(N_X(L)). \label{eq:sandwich-on-kappa}
\end{equation}
We apply \Cref{lemma:expectation-of-overshoot} with $R_i=V_{X,i}$, $W_i=V_{Y,i}$, and $m=L$.
The assumptions hold because $V_{X,i} \geq 1$ under $\bfx$-first, and $V_{X,i},V_{Y,i} \leq T_i$, while \Cref{lemma:T-has-exponential-tail} implies that $T_i$ has finite second moment. 
Therefore,
$$\mathbb{E}[Y_{\mathrm{tot}}(N_X(L))] = \frac{\mathbb{E}[V_Y]}{\mathbb{E}[V_X]}L+O_q(1),$$
and
$$\mathbb{E}[Y_{\mathrm{tot}}(N_X(L)-1)] = \frac{\mathbb{E}[V_Y]}{\mathbb{E}[V_X]}L+O_q(1).$$
Together with \Cref{eq:sandwich-on-kappa}, this gives
$$\mathbb{E}[\kappa] = \frac{\mathbb{E}[V_Y]}{\mathbb{E}[V_X]}L+O_q(1). $$
Intuitively, this result is what we expect, as we expect roughly $L/\mathbb{E}[V_X]$ full rotations to finish $\bfx$, and $\bfy$ will synthesize roughly $\mathbb{E}[V_Y]$ symbols during this period.
Next, by \Cref{claim:E[X]-and-E[Y]-final},
$$\frac{\mathbb{E}[V_Y]}{\mathbb{E}[V_X]} = \frac{q-1}{q+3}, $$
and hence
\begin{equation}
    \mathbb{E}[\kappa] = \frac{q-1}{q+3}L+O_q(1). \label{eq:expected-kappa}
\end{equation}
Under $\bfx$-first, the strand $\bfx$ evolves exactly as a single strand synthesized under the periodic sequence.
Thus, if $T_x$ is the time at which $\bfx$ is completed, then
\begin{equation}
    \mathbb{E}[T_x]=\frac{q+1}{2}L. \label{eq:expected-T_x}
\end{equation}
After time $T_x$, only the remaining suffix of $\bfy$ must be synthesized.
This suffix has length $L-\kappa$, and its symbols are still i.i.d. uniform. Therefore, its expected solo synthesis time is $\frac{q+1}{2}(L-\kappa)$.
Consequently,
$$\mathbb{E}[T_{\bfx\text{-first}}(\bfx,\bfy)] = \mathbb{E}[T_x] + \frac{q+1}{2}\mathbb{E}[L-\kappa].$$
Substituting \Cref{eq:expected-kappa} and \Cref{eq:expected-T_x}, we obtain
\begin{align*}
    \mathbb{E}[T_{\bfx\text{-first}}(\bfx,\bfy)] &= \frac{q+1}{2}L + \frac{q+1}{2} \left( L-\frac{q-1}{q+3}L+O_q(1) \right) \\
    &= \frac{q+1}{2}L \left( 2-\frac{q-1}{q+3} \right) +O_q(1) \\
    &= \left(\frac{q+3}{2} + \frac{q-1}{q+3}\right)L + O_q(1).
\end{align*}
\end{IEEEproof}

\begin{lemma}\label{lemma:expectation-of-overshoot}
    Let $(R_i,W_i)_{i\geq 1}$ be i.i.d. random integer pairs, where $W_i \geq 0$, $R_i\geq 1$, and $\mathbb{E}[(R_i+W_i)^2]<\infty$.
    Define
    $$R_{\mathrm{tot}}(n) \coloneqq \sum_{i=1}^n R_i,\qquad W_{\mathrm{tot}}(n) \coloneqq \sum_{i=1}^n W_i,$$
    with $R_{\mathrm{tot}}(0)=W_{\mathrm{tot}}(0)=0$ by convention.
    Then, for every integer $m \geq 1$, define $N_m \coloneqq \min\{n:R_{\mathrm{tot}}(n)\ge m\}$.
    Thus, the following holds:
    \begin{align*}
        \mathbb{E}[W_{\mathrm{tot}}(N_m)] &= \frac{\mathbb{E}[W_1]}{\mathbb{E}[R_1]}m+O(1), \\
        \mathbb{E}[W_{\mathrm{tot}}(N_m-1)] &= \frac{\mathbb{E}[W_1]}{\mathbb{E}[R_1]}m+O(1).
    \end{align*}
\end{lemma}
\begin{IEEEproof}
    Let $\mu_R \coloneqq \mathbb{E}[R_1]$ and $\mu_W \coloneqq \mathbb{E}[W_1]$.
    Then, since $R_i\ge 1$, we have $N_m \leq m$, so $N_m$ is a bounded stopping time with respect to the pairs $(R_i,W_i)$.
    The processes $R_{\mathrm{tot}}(n)-n\mu_R$ and $W_{\mathrm{tot}}(n)-n\mu_W$ are martingales.
    Therefore, by optional stopping~\cite[Theorem 3.6]{Bhattacharya2007-ca},
    $$0 = \mathbb{E}[R_{\mathrm{tot}}(N_m)-N_m\mu_R] = \mathbb{E}[R_{\mathrm{tot}}(N_m)]-\mu_R\mathbb{E}[N_m]$$
    Thus, 
    $$\mathbb{E}[R_{\mathrm{tot}}(N_m)]=\mu_R\mathbb{E}[N_m],$$
    and similarly 
    $$\mathbb{E}[W_{\mathrm{tot}}(N_m)]=\mu_W\mathbb{E}[N_m].$$
    Let $U_m:=R_{\mathrm{tot}}(N_m)-m$ be the overshoot. By Lorden's inequality~\cite{Lorden1970}, since $\mathbb{E}[R_1^2]<\infty$,
    $$\sup_{m\ge 1}\mathbb{E}[U_m]<\infty.$$
    Hence
    $$\mathbb{E}[N_m] = \frac{\mathbb{E}[R_{\mathrm{tot}}(N_m)]}{\mu_R} = \frac{m+\mathbb{E}[U_m]}{\mu_R} = \frac{m}{\mu_R}+O(1).$$
    Thus,
    $$\mathbb{E}[W_{\mathrm{tot}}(N_m)] = \frac{\mu_W}{\mu_R}m+O(1).$$
    Since $W_{\mathrm{tot}}(N_m)-W_{\mathrm{tot}}(N_m-1)=W_{N_m}$, it is enough to show that $\mathbb{E}[W_{N_m}]=O(1)$. 
    Indeed, let 
    $$\cC_n^{(m)} \coloneqq \{R_{\mathrm{tot}}(n-1)<m\le R_{\mathrm{tot}}(n-1)+R_n\},$$ be the event that the cumulative $R$-process crosses level $m$ during the $n$-th renewal block.
    $$\mathbb{E}[W_{N_m}] = \sum_{n\ge 1} \mathbb{E}\!\left[ W_n \bfone_{\cC_n^{(m)}} \right].$$
    Now, for a fixed $n$, decompose according to the value of $R_{\mathrm{tot}}(n-1)$.
    Since $\mathcal{C}_n^{(m)}$ can occur only when $R_{\mathrm{tot}}(n-1)\in\{0,1,\ldots,m-1\}$, we have
    $$\bfone_{\mathcal{C}_n^{(m)}} = \sum_{s=0}^{m-1} \bfone_{\{R_{\mathrm{tot}}(n-1)=s\}} \bfone_{\{R_n\ge m-s\}}.$$
    Therefore,
    $$\mathbb{E}\left[ W_n\bfone_{\mathcal{C}_n^{(m)}} \right] = \sum_{s=0}^{m-1} \mathbb{E}\left[ W_n \bfone_{\{R_{\mathrm{tot}}(n-1)=s\}} \bfone_{\{R_n\ge m-s\}} \right].$$
    The event $\{R_{\mathrm{tot}}(n-1)=s\}$ depends only on the first $n-1$ vectors, while $(R_n,W_n) \sim (R_1,W_1)$ is independent of them. Thus,
    \begin{align*}
        &\mathbb{E}\left[ W_n \bfone_{\{R_{\mathrm{tot}}(n-1)=s\}} \bfone_{\{R_n\ge m-s\}} \right] \\
        &= \Pr(R_{\mathrm{tot}}(n-1)=s) \mathbb{E}\left[ W_1\bfone_{\{R_1\ge m-s\}} \right].
    \end{align*}
    Substituting this into the previous expression gives
    $$\mathbb{E}[W_{N_m}] = \sum_{n\ge 1}\sum_{s=0}^{m-1} \Pr(R_{\mathrm{tot}}(n-1)=s) \mathbb{E}\left[ W_1\bfone_{\{R_1\ge m-s\}} \right].$$
    Since all terms in the double sum are nonnegative, Tonelli's theorem allows to interchange the order of summation. Hence,
    \begin{align*}
        \mathbb{E}[W_{N_m}]
        \!=\! \sum_{s=0}^{m-1} \!\left(\!\sum_{n\geq 1} \Pr(R_{\mathrm{tot}}(n\!-\!1)\!=\!s) \!\right) \!\mathbb{E}\!\left[ W_1\bfone_{\{R_1\ge m-s\}} \right]\!.
    \end{align*}
    Since $R_i\geq 1$, the process $R_{\mathrm{tot}}(n)$ is strictly increasing. Therefore, for each fixed $s$,
    $$\sum_{n\ge 1}\Pr(R_{\mathrm{tot}}(n-1)=s)\le 1.$$
    Hence,
    $$\mathbb{E}[W_{N_m}] \leq \sum_{s=0}^{m-1} \mathbb{E}\left[ W_1\bfone_{\{R_1\ge m-s\}} \right].$$
    Changing variables $r=m-s$, we obtain 
    $$\mathbb{E}[W_{N_m}] \leq \sum_{r=1}^{m} \mathbb{E}\left[ W_1\bfone_{\{R_1\ge r\}} \right].$$
    Finally, since $\sum_{r=1}^{m} \bfone_{\{R_1 \geq r\}} = \min\{m,R_1\} \leq R_1$, and $W_1,R_1 \geq 0$,
    $$\sum_{r=1}^{m} \mathbb{E}\left[ W_1\bfone_{\{R_1\ge r\}}\right] = \mathbb{E}\left[ W_1\sum_{r=1}^{m}\bfone_{\{R_1\ge r\}} \right] \leq \mathbb{E}[W_1R_1].$$
    The assumption $\mathbb{E}[(R_1+W_1)^2]<\infty$ implies
    $\mathbb{E}[W_1R_1]<\infty$. 
    Therefore, $\mathbb{E}[W_{N_m}]=O(1)$, as required.
\end{IEEEproof}

\claimSNtwoL*
\begin{IEEEproof}
    Let $\mathcal{F}_{i-1}$ denote the history up to the beginning of the $i$-th full rotation.
    Under any policy that resolves ties using only past information, the decision at the initial tie of the $i$-th full rotation is $\mathcal{F}_{i-1}$-measurable. Conditional on this decision, the future unread symbols are independent of $\mathcal{F}_{i-1}$.
    If the tie is resolved in favor of $\bfx$, the full rotation has the $\bfx$-first distribution; if it is resolved in favor of $\bfy$, the same distribution is obtained after swapping the two coordinates.
    In either case, the pair $(Z_i,T_i)$ has the same distribution and is independent of $\mathcal{F}_{i-1}$. Therefore, the pairs $(Z_i,T_i)_{i\geq 1}$, are i.i.d.
    By \Cref{claim:E[X]-and-E[Y]-final},
    $$\mathbb{E}[Z_i] = \mathbb{E}[V_X]+\mathbb{E}[V_Y] = q, \qquad \mathbb{E}[T_i] = \frac{q(q+3)}{4}.$$
    Moreover, $Z_i\geq 1$ since every full rotation contains the initial advance, and $Z_i\leq T_i$.
    By \Cref{lemma:T-has-exponential-tail}, $T_i$ has a finite second moment, and hence $\mathbb{E}[(Z_i+T_i)^2]<\infty$.
    Applying \Cref{lemma:expectation-of-overshoot} with $R_i=Z_i$, $W_i=T_i$, and $m=2L$, we obtain
    \begin{align*}
        \mathbb{E}[S_{N_{2L}}] &= \mathbb{E}[S_{N_{2L}-1}] = \frac{\mathbb{E}[T_i]}{\mathbb{E}[Z_i]}\cdot 2L+O_q(1) \\
        &= \frac{q(q+3)/4}{q}\cdot 2L+O_q(1) = \frac{q+3}{2}L+O_q(1).
    \end{align*}
\end{IEEEproof}

\claimdriftimbalance*
\begin{IEEEproof}
    The proof relies on the excursion decomposition of the imbalance process under the LF policy.
    By \Cref{lemma:drift-outside-Rq-is-low-prob}, there exist constants
    $\theta>0$, $R_q<\infty$, and $\rho_q<1$ such that the imbalance process satisfies the geometric drift bound \cite[Section 15]{Meyn2009-dp}:
    $
    \mathbb E[e^{\theta |d_n|}\mid d_{n-1}=d] \leq \rho_q e^{\theta |d|},
    $
    for $\left|d\right|>R_q$.
    Let 
    $C_q\coloneqq\{d\in\mathbb{Z} \colon \left|d\right|\leq R_q\}.$
    Next we decompose the process into excursions between
    successive returns to $C_q$. 
    If $H_k$ is the total progress during the $k$-th excursion and $M_k$ is the maximum
    value of $\left|d_n\right|$ during that excursion,
    i.e., 
    $$M_k=\max_{\tau_{k-1} \leq i < \tau_{k}} |d_i|, \qquad H_k = A_{\tau_k} - A_{\tau_{k-1}},$$
    then \Cref{lemma:finite-moment-of-Hk-Mk} shows that for every $k$,
    $$\sup_{d\in C_q} \mathbb{E} [M_k H_k \mid d_{\tau_{k-1}}=d]<\infty.$$
    Fix $m \geq 1$, and define 
    $$J_m \coloneqq \min\{k \colon A_{\tau_k}\geq m\},$$
    the index of the excursion during which the cumulative progress first reaches or exceeds $m$.
    Since $N_m$ occurs during this excursion, we have $\left|d_{N_m}\right| \leq M_{J_m}$.
    By \Cref{lemma:expected-drift-at-stopping-is-const}, 
    $$\sup_{m \geq 1}\mathbb{E} [M_{J_m}]<\infty.$$
    Therefore,
    $$\sup_{m\geq1}\mathbb E[\left|d_{N_m}\right|] \leq \sup_{m \geq 1}\mathbb E[M_{J_m}]<\infty.$$
    By taking $m=2L$, we conclude the proof.
\end{IEEEproof}

\claimextratime*
\begin{IEEEproof}
    Define the remaining number of symbols needed for the completion of the lagging strand at $N_{2L}$ as 
    $$
        R_L \coloneqq \max \left\{ 0, L - \min \left\{ X_{\text{tot}}(N_{2L}), Y_{\text{tot}}(N_{2L}) \right\} \right\}.
    $$
    Since $X_{\text{tot}}(n) = (A_n + d_n)/2$, and $Y_{\text{tot}}(n) = (A_n - d_n)/2$, 
    $$
        \min \left\{ X_{\text{tot}}(N_{2L}), Y_{\text{tot}}(N_{2L}) \right\} = \frac{A_{N_{2L}}-|d_{N_{2L}}|}{2},
    $$
    and since $A_{N_{2L}} \geq 2L$, we get $R_L \leq \frac{|d_{N_{2L}}|}{2}$, and by \Cref{claim:imbalance-drift-bound},
    \begin{equation}
        \mathbb{E}[R_L] \leq \frac{1}{2}\mathbb{E}[|d_{N_{2L}}|] = O_q(1). \label{eq:R_L-is-const}
    \end{equation}
    Suppose w.l.o.g. that $\bfx$ is ahead at full rotation $N_{2L}$.
    If $R_L=0$, we are done. 
    Otherwise, $R_L>0$ and $\bfy$ needs to advance $R_L$ more symbols.
    Then, each full rotation between $N_{2L}$ until $N_{*}$ starts with $\bfy$-advance, since it is the laggard, thus $Y_{\text{tot}}$ is increased by at least one.
    Therefore, $N_{*} - N_{2L} \leq R_L$.
    The extra number of slots after the $N_{2L}$-th full rotation ends and until the end of the $N_{*}$-th full rotation is:
    $$S_{N_{*}} - S_{N_{2L}} = \sum_{k=N_{2L}+1}^{N_{*}} T_k.$$
    Since $R_L$ is the number of symbols still required by the lagging strand, and under LF each subsequent full rotation advances the lagging strand by at least one symbol, at most $R_L$ additional full rotations are needed. 
    Let $c_q$ denote the maximum expected length of a full rotation, which is finite by \Cref{lemma:T-has-exponential-tail}.
    Then, conditioned on the state of the process at time $N_{2L}$, the expected remaining synthesis time is at most $c_q \mathbb{E}[R_L]$.
    Using \Cref{eq:R_L-is-const}, we conclude the proof.
\end{IEEEproof}

\begin{lemma}\label{lemma:Z-and-Delta-has-exponential-tail}
    For all $i\geq 1$, let $Z_i \coloneqq V_{X,i} + V{Y,i}$ and $\Delta_i \coloneqq V_{X,i} - V{Y,i}$.
    Let $\theta_q>0$ be the constant for which $\mathbb{E}[e^{\theta_q T}] < \infty$ by  \Cref{lemma:T-has-exponential-tail}.
    Then, for all $|\theta| \leq \theta_q$:
    \begin{align*}
        &\sup_{d}\mathbb{E}[e^{\theta Z_i} \mid d_{i-1} = d] < \infty, \\
        &\sup_{d}\mathbb{E}[e^{\theta \left| \Delta_{i} \right|} \mid d_{i-1} = d] < \infty.
    \end{align*}
    That is, $Z_i$ and $\Delta_i$ have a finite exponential moment.
\end{lemma}
\begin{IEEEproof}
    During one full rotation, at most one strand advances in each slot. 
    Therefore, $1 \leq Z_i \leq T_i$.
    Moreover, $0 \leq \left| \Delta_{i} \right| \leq X_i+Y_i = Z_i \leq T_i$.
    Thus, by \Cref{lemma:T-has-exponential-tail} we conclude the lemma.
\end{IEEEproof}

\begin{lemma}\label{lemma:drift-outside-Rq-is-low-prob}
    There exist constants $\theta>0$, $R_q<\infty$, and $\rho_q<1$ such that:
    \begin{align*}
        \mathbb{E}[e^{\theta \left| d_i \right|} \mid d_{i-1}=d] \leq \rho_q e^{\theta|d|} ,
    \end{align*}
    for all $|d|>R_q$.
\end{lemma}
\begin{IEEEproof}
    Let $0 < \theta \leq \theta_q$, where $\theta_q$ is from \Cref{lemma:T-has-exponential-tail}. We will later choose a specific $\theta$ small enough. 
    Define $V(d) = e^{\theta \left|d\right|}$.
    Assume first that $d>0$ is fixed. Thus, under LF, the next full rotation is $\bfy$-first.
    By \Cref{claim:E[X]-E[Y]} and symmetry,
    \begin{align*}
	   \mathbb{E}[\Delta_{i} \mid d_{i-1}=d] = -\delta_q, \quad \delta_q = \frac{2q}{q+1}>0.
    \end{align*}
    Now,
    $$
	   \frac{\mathbb{E}[V(d+ \Delta_{i}) \mid d_{i-1}=d]}{V(d)}=\mathbb{E}\left[e^{\theta \left( \left| d+ \Delta_{i} \right| -d \right)} \mid d_{i-1}=d\right].
    $$
	If $d+\Delta_i \geq 0$, then $\left|d+\Delta_i\right|-d=\Delta_i$.
	If $d+\Delta_i < 0$, then the jump overshoots zero, and $\left|d+\Delta_i\right|-d = -2d-\Delta_i$.
	Therefore,
	$$
		e^{\theta \left( \left| d+ \Delta_{i} \right| -d \right)} \leq e^{\theta \Delta_{i}} + e^{-\theta \Delta_{i}} \bfone_{\{\Delta_i<-d\}},
	$$
    and taking expectation gives
    \begin{align*}
        \frac{\mathbb{E}[V(d+ \Delta_{i}) \mid d_{i-1}\!=\!d]}{V(d)} &\leq \mathbb{E}\left[ e^{\theta \Delta_{i}} \mid d_{i-1}\!=\!d \right] \\
        &+ \mathbb{E}\left[ e^{-\theta \Delta_{i}} \bfone_{\{\Delta_i<-d\}} \mid d_{i-1}\!=\!d \right].
    \end{align*}
    Next, since $\mathbb{E}[\Delta_{i} \mid d_{i-1}=d] = -\delta_q$, and since \Cref{lemma:Z-and-Delta-has-exponential-tail} gives an exponential moment for all $\left|\theta\right| < \theta_q$, for sufficiently small $\theta > 0$, the Taylor expansion around $\theta=0$ is:
    \begin{equation}
        \mathbb{E}\left[ e^{\theta \Delta_{i}} \mid d_{i-1}=d \right] \leq 1 - \frac{\theta\delta_q}{2}. \label{eq:bound-drift-1}
    \end{equation}
    This bounds the first term. For the second term, note that on the event $\{\Delta_i<-d\}$, we have $\left|\Delta_i\right|>d$ and $e^{\theta\Delta_i} = e^{\theta\left|\Delta_i\right|}$.
    Also, $e^{\theta\left|\Delta_i\right|} = e^{\theta_q\left|\Delta_i\right|}e^{-(\theta_q-\theta)\left|\Delta_i\right|} \leq e^{\theta_q\left|\Delta_i\right|}e^{-(\theta_q-\theta)d}$. 
    Therefore, $e^{-\theta \left|\Delta_i\right|} \bfone_{\{\left|\Delta_i\right|>d\}} \leq e^{\theta_q\left|\Delta_i\right|}e^{-(\theta_q-\theta)d}$, and thus,
    \footnotesize\begin{align*}
        \mathbb{E}\left[ e^{-\theta \Delta_{i}} \bfone_{\{\Delta_i<-d\}} \mid d_{i-1}=d \right] &\leq \mathbb{E}\left[ e^{\theta \left|\Delta_i\right|} \bfone_{\{\left|\Delta_i\right|>d\}} \mid d_{i-1}=d \right] \\
        &\leq e^{-(\theta_q-\theta)d}\mathbb{E}\left[ e^{\theta_q\left|\Delta_i\right|} \mid d_{i-1}=d \right].
    \end{align*}
    \normalsize By \Cref{lemma:Z-and-Delta-has-exponential-tail} the second term is finite.
    Also, $e^{-(\theta_q-\theta)d} \to 0$ when $d\to\infty$, thus
    $$\lim_{d\to\infty} \mathbb{E}\left[ e^{-\theta \Delta_{i}} \bfone_{\{\Delta_i<-d\}} \mid d_{i-1}=d \right] = 0,$$
    and there exist $R_q$ such that for all $d>R_q$,
    \begin{equation}
        \mathbb{E}\left[ e^{-\theta \Delta_{i}} \bfone_{\{\Delta_i<-d\}} \mid d_{i-1}=d \right] \leq \frac{\theta\delta_q}{4}. \label{eq:bound-drift-2}
    \end{equation}
    By \Cref{eq:bound-drift-1} and \Cref{eq:bound-drift-2}, for all $d>R_q$,
    \begin{align*}
        \frac{\mathbb{E}[V(d+ \Delta_{i}) \mid d_{i-1}\!=\!d]}{V(d)} \leq 1-\frac{\theta\delta_q}{4}.
    \end{align*}
    The case $d<0$ is symmetric. Thus, for all $|d| > R_q$, 
    \begin{align*}
        \mathbb{E}[V(d + \Delta_{i}) \mid d_{i-1}\!=\!d] \leq \rho_q V(d),
    \end{align*}
    where $\rho_q = 1-\frac{\theta\delta_q}{4} < 1$, and, by definition, $V(d) = e^{\theta|d|}$ and $d_i = d + \Delta_{i}$.
\end{IEEEproof}

\begin{lemma}\label{lemma:finite-moment-of-Hk-Mk}
    Let $C_q \coloneqq \{d \in \mathbb{Z} \colon \left|d\right| \leq R_q\}$.
    Denote $\tau_0=0$ and $\tau_{k+1} = \min\{i>\tau_k \colon d_i \in C_q\}$ as returns (in the end of a full rotation) to a bounded interval of the drifts. 
    The interval of full rotations $[\tau_{k-1}, \tau_{k})$ is called the $k$-th excursion (away from $C_q$).
    That is, each excursion begins when the imbalance process enters $C_q$ and ends at the next return to $C_q$.
    Let 
    $$H_k = A_{\tau_k} - A_{\tau_{k-1}},$$
    be the total progress during the $k$-th excursion, and let 
    $$M_k=\max_{\tau_{k-1} \leq i < \tau_{k}} |d_i|,$$
    be the maximum imbalance during the $k$-th excursion.
    Thus, for every $k \in \mathbb{N}$, 
    \begin{align*}
        \sup_{d\in C_q}\mathbb{E}[M_k H_k \mid d_{\tau_{k-1}} = d] < \infty.
    \end{align*}
\end{lemma}
\begin{IEEEproof}
    First, by \Cref{lemma:drift-outside-Rq-is-low-prob}, if $d \notin C_q$:
    \begin{equation}
       \mathbb{E}[V(d_i) \mid d_{i-1}=d] \leq \rho_q V(d), \quad V(d) \coloneqq e^{\theta \left| d \right|}. \label{eq:drift-from-lemma-outside-Rq}
    \end{equation}
    Moreover, since $C_{q}$ is finite and $\Delta_i$ has a finite exponential moment by \Cref{lemma:Z-and-Delta-has-exponential-tail}, if $d\in C_q$:
    \begin{equation}
       K_q \coloneqq \sup_{d\in C_q} \mathbb{E}[V(d_i) \mid d_{i-1}=d] < \infty. \label{eq:drift-inside-Rq}
    \end{equation}
    We prove the bound for one generic excursion started from an arbitrary state $d \in C_q$.
    Fix $k$ to be the $k$-th excursion. Then, iterating \Cref{eq:drift-from-lemma-outside-Rq}, with the base case being \Cref{eq:drift-inside-Rq}:
    $$\sup_{d \in C_q} \mathbb{E}\left[ V(d_n)\bfone_{\{\tau_{k}>n \}} \mid d_{\tau_{k-1}} = d\right] \leq K_q \rho_q^{n-\tau_{k-1}-1},$$
    where $n>\tau_{k-1}$. Since $V(d_n) \geq 1$, for all $n>\tau_{k-1}$:
    $$\sup_{d \in C_q} \Pr\left( \tau_{k}>n \mid d_{\tau_{k-1}} = d \right) \leq K_q \rho_q^{n-\tau_{k-1}-1}.$$
    Thus, $\tau_{k}$ has a geometric tail, uniformly over all starting states $d \in C_q$.
    Next, denote $\tau \coloneqq \tau_{k}-\tau_{k-1}$ and thus, 
    \begin{equation}
        \sup_{d \in C_q} \mathbb{E}[\tau^2 \mid \tau_{k-1}=d] < \infty. \label{eq:finite-moment-of-excursion-time}
    \end{equation}
    Note that $\tau$ has a finite exponential moment, but we only need the finiteness of the second moment. 
    Since $H_k=\sum_{i=\tau_{k-1}}^{\tau_{k}-1} Z_i$, using Cauchy-Schwarz, 
    $$H_k^2 \leq \tau \sum_{i=\tau_{k-1}}^{\tau_{k}-1} (Z_i)^2.$$
    Denote $\mathbb{E}_{d}[\bullet] \coloneqq \mathbb{E}[\bullet \mid \tau_{k-1}=d]$. 
    By Hölder's inequality:
    \begin{align*}
        &\mathbb{E}_{d}[H_k^2] \leq \sum_{i \geq 1} \mathbb{E}_{d}[\tau (Z_i)^2 \bfone_{\{\tau_{k-1} \leq i < \tau_k\}}] \\
        &\leq \sum_{i \geq 1} (\mathbb{E}_{d}[\tau^2])^{1/2} \cdot (\mathbb{E}_{d}[(Z_i)^4])^{1/2} \bfone_{\{i\leq\tau\}} \\
        &\cdot \sum_{i \geq 1}\left(\Pr\left( \tau_{k-1} \leq i < \tau_k \mid d_{\tau_{k-1}} = d \right)\right)^{1/2}.
    \end{align*}
    And by \Cref{lemma:Z-and-Delta-has-exponential-tail} and \Cref{eq:finite-moment-of-excursion-time}, one concludes that 
    $$\sup_{d \in C_q} \mathbb{E}_{d}[H_k^2]<\infty.$$
    Similarly, since $M_k \leq R_q \sum_{i=\tau_{k-1}}^{\tau_{k}-1} \left|\Delta_i\right|$, it holds that $$\sup_{d \in C_q} \mathbb{E}_{d}[M_k^2]<\infty.$$
    Finally, by Cauchy-Schwarz,
    $$\sup_{d \in C_q} \mathbb{E}_{d}[M_k H_k] \leq \sup_{d \in C_q} \sqrt{\mathbb{E}_{d}[M_k^2]\mathbb{E}_{d}[H_k^2]} < \infty.$$
\end{IEEEproof}

\begin{lemma}\label{lemma:expected-drift-at-stopping-is-const}
    Let $J_m \coloneqq \min\{ k \colon A_{\tau_k} \geq m\}$ be the index of the excursion during which the cumulative progress first reaches or exceeds $m$.
    Then, for a stopping time $N_m$ and a fixed $q$, it holds that $\left|d_{N_{m}}\right| \leq M_{J_m}$. Thus,
    \begin{align*}
        \mathbb{E}[|d_{N_{2L}}|] \leq \sup_{m}\mathbb{E}[|d_{N_{m}}|] \leq \sup_{m}\mathbb{E}[M_{J_m}] < \infty.
    \end{align*}
\end{lemma}
\begin{IEEEproof}
    $N_m$ occurs during the $J_m$-th excursion.
    Hence, $\left|d_{N_{m}}\right| \leq M_{J_m}$.
    It remains to prove that $\sup_{m}\mathbb{E}[M_{J_m}] < \infty$.
    Let $S_k \coloneqq A_{\tau_k}$.
    So $S_k$ is the cumulative progress at the end of the $k$-th excursion, and $H_k=S_k-S{k-1}$.
    For fixed $m$, $J_m=k$ exactly when $S_{k-1} < m \leq S_k$.
    Now:
    $$ \mathbb{E}[M_{J_m}]= \sum_{k=1}^{\infty} \mathbb{E}\left[M_k \bfone_{\{S_{k-1} < m \leq S_{k}\}}\right]. $$
    Next, condition on the past up to the start of excursion $k$. 
    If $S_{k-1}=s<m$, then for excursion $k$ to contain the cumulative progress value $m$, we must have $H_k \geq m-s$.
    Therefore,
    \footnotesize$$ \mathbb{E}\left[M_k \bfone_{\{S_{k-1} < m \leq S_{k}\}} \mid S_{k-1}=s\right] \leq \sup_{d\in C_q} \mathbb{E}_d\left[M_k \bfone_{\{H_k \geq m-s\}} \right]. $$
    \normalsize Hence,
    $$ \mathbb{E}[M_{J_m}] \leq \sum_{s=0}^{m-1} \sup_{d\in C_q} \mathbb{E}_d\left[M_k \bfone_{\{H_k \geq m-s\}} \right].$$
    Setting $r=m-s$ we get:
    $$ \mathbb{E}[M_{J_m}] \leq \sum_{r=1}^{m} \sup_{d\in C_q} \mathbb{E}_d\left[M_k \bfone_{\{H_k \geq r\}} \right].$$
    Since $C_q$ is finite,
    $$ \mathbb{E}[M_{J_m}] \leq \sum_{d\in C_q}\sum_{r=1}^{\infty} \mathbb{E}_d\left[M_k \bfone_{\{H_k \geq r\}} \right].$$
    Finally, $\sum_{r=1}^{\infty}\bfone_{\{H_k \geq r\}}=H_k$, so by monotone convergence,
    $$ \sum_{r=1}^{\infty} \mathbb{E}_d\left[M_k \bfone_{\{H_k \geq r\}} \right] = \mathbb{E}_d\left[M_k H_k \right].$$
    Thus, by \Cref{lemma:finite-moment-of-Hk-Mk}, 
    $$ \mathbb{E}[M_{J_m}] \leq \sum_{d\in C_q} \mathbb{E}_d\left[M_k H_k \right] = O_q(1).$$
\end{IEEEproof}

\begin{lemma}\label{lemma:LF1-bounded-imbalance}
Under the \(\mathrm{LF}_1\) policy in the infinite i.i.d. binary process,
let $D_t \coloneqq i_t-j_t$ and $\tau_m \coloneqq \min\{t:G_t\ge m\}$.
Then,
$$\sup_{m\ge 0}\mathbb E\!\left[|D_{\tau_m}|\right]<\infty.$$
\end{lemma}
\begin{IEEEproof}
Let $W_m \coloneqq D_{\tau_m}$ be the imbalance after the $m$-th synthesized symbol in the infinite process.
Since exactly one strand advances at each progress epoch,
$$W_{m+1}-W_m\in\{-1,+1\}.$$
We first prove a finite-state drift estimate for an auxiliary rule. 
Consider the fixed rule $\Pi_y$ that agrees with $\mathrm{LF}_1$, except that whenever the look-ahead symbols are equal, it always advances $\bfy$.

For a state $s=(a,b,c,d)\in\{0,1\}^4$, let $N_x^{(r)}(s)$ and $N_y^{(r)}(s)$ be the number of advances of $\bfx$ and $\bfy$, respectively, during the next $r$ progress epochs (i.e., a time slot in which one strand advances). Define
$$F_r(s) \coloneqq \mathbb E\left[N_x^{(r)}(s)-N_y^{(r)}(s)\right].$$
For $r \geq 0$, the values $F_{r+1}$ are obtained from $F_r$ by conditioning on the first progress epoch.
We write $\bar{z} \coloneqq 1-z$, and we display the representative cases.
First, if only $\bfx$ can advance, namely $s=(0,1,c,d)$, then the first progress epoch contributes $+1$ to the imbalance.
After this advance, the next state is $(\bar c,0,U,\bar d)$, where $U\sim\operatorname{Bernoulli}(1/2)$ is the newly exposed look-ahead bit of $\bfx$.
Hence
$$ F_{r+1}(0,1,c,d) = 1+\frac{1}{2}F_r(\bar{c},0,0,\bar{d}) +\frac{1}{2}F_r(\bar{c},0,1,\bar{d}).$$
The only tie state in which the look-ahead forces an advance of \(\bfx\)
under $\Pi_y$ is $(0,0,1,0)$. 
Therefore,
$$F_{r+1}(0,0,1,0) = 1+\frac12F_r(0,1,0,1) + \frac12F_r(0,1,1,1).$$
In every other tie state $(0,0,c,d)\neq(0,0,1,0)$, the rule $\Pi_y$ advances $\bfy$.
Thus, the first progress epoch contributes $-1$, and
$$F_{r+1}(0,0,c,d) = -1+\frac12F_r(1,\bar d,\bar c,0) +\frac12F_r(1,\bar d,\bar c,1).$$
Finally, if $a=b=1$, no strand advances in the current synthesis slot.
Since $F_r$ counts progress epochs rather than synthesis slots, $F_{r+1}(1,1,c,d)=F_{r+1}(0,0,\bar c,\bar d)$.
The remaining case, in which only $\bfy$ can advance, is obtained in the same way as the first displayed equation, with first increment $-1$ and with the newly exposed look-ahead bit belonging to $\bfy$.

Starting from $F_0 = 0$, these equations determine $F_r(s)$ for all $s\in\{0,1\}^4$ and all $r$.
Evaluating this finite recursion gives
$$\begin{array}{c|ccccccccc}
r        &1&2&3&4&5&6&7&8&9\\ \hline \\[-8pt]
\max_s F_r(s)
&1&2&\frac32&2&\frac54&\frac{19}{16}
&\frac{19}{32}&\frac{11}{32}&-\frac{5}{32}.
\end{array}$$
Thus, $r=9$ is the first block length for which the expected drift is uniformly negative over all starting states. In particular, 
\begin{equation}
    F_9(s) \leq -\frac{5}{32} \quad\text{for every }s\in\{0,1\}^4. \label{eq:F_9(s)}
\end{equation}

We now explain why this finite-state computation applies to the actual
\(\mathrm{LF}_1\) policy when the imbalance is large. Let \(\mathcal H_m\)
be the history up to time \(\tau_m\). If \(W_m\ge 9\), then during the next
nine progress epochs the imbalance remains positive before every decision,
because each progress epoch changes \(W_m\) by at most one. Hence, whenever
the look-ahead symbols are equal during these nine progress epochs, the
actual \(\mathrm{LF}_1\) policy advances \(\bfy\). Therefore, conditioned
on \(\mathcal H_m\), the next nine progress epochs under the actual
\(\mathrm{LF}_1\) policy have the same distribution as under the fixed rule
\(\Pi_y\), started from the current state \(S_{\tau_m}\). 
By \Cref{eq:F_9(s)}:
\begin{equation}
    \mathbb E[W_{m+9}-W_m\mid \mathcal H_m] \leq -\frac{5}{32}, \qquad W_m\ge 9. \label{eq:E-imbalance-in-9-steps-y}
\end{equation}
By symmetry, if \(W_m\le -9\), then during the next nine progress epochs
the actual policy agrees with the fixed rule \(\Pi_x\) that always advances
\(\bfx\) in unresolved look-ahead ties. Hence,
\begin{equation}
    \mathbb E[W_{m+9}-W_m\mid \mathcal H_m] \ge \frac{5}{32}, \qquad W_m\le -9. \label{eq:E-imbalance-in-9-steps-x}
\end{equation}
We now convert the block drift into a exponential Lyapunov function.
Let 
$$V_m:=e^{\theta \left|W_m\right|}.$$
Suppose first that $W_m\ge 9$, and set $\Delta_m \coloneqq W_{m+9}-W_m$.
Hence, by \Cref{eq:E-imbalance-in-9-steps-y},
$$\mathbb E[\Delta_m\mid\mathcal H_m]\le -\delta,$$
where $\delta \coloneqq \frac{5}{32}$.
Moreover, $V_{m+9}=V_m e^{\theta\Delta_m}$.
Since $\Delta_m$ is bounded and has conditional expectation at most $-\delta$, similarly to \Cref{eq:bound-drift-1}, choosing $\theta>0$ sufficiently small gives a constant $\rho<1$ such that 
$$\mathbb E[e^{\theta\Delta_m}\mid\mathcal H_m]\leq \rho.$$
Therefore, since $V_m$ is $\cH$-measurable,
$$\mathbb E[V_{m+9}\mid\mathcal H_m]\le \rho V_m, \qquad W_m\ge 9.$$
The case $W_m \leq -9$ is symmetric: applying the same argument to $-\Delta_m$ gives the same bound. Thus,
$$\mathbb E[V_{m+9}\mid\mathcal H_m]\le \rho V_m, \qquad |W_m|\ge 9.$$
On the other hand, if $|W_m|<9$, then after nine progress epochs $|W_{m+9}|\leq 18$, and therefore $V_{m+9}\leq e^{18\theta}$.
Thus, for every $m$,
$$\mathbb E[V_{m+9}\mid \mathcal H_m] \leq \rho V_m+e^{18\theta}.$$

Next, we apply this along the subsequence $m=9n$, $n \geq 0$. 
Let $A_n:=\mathbb E[V_{9n}]$.
Then, by the law of total expectation,
$$A_{n+1} = \mathbb{E}\left[\mathbb E[V_{9n+9}\mid\mathcal H_{9n}]\right] \leq \rho A_n+e^{18\theta}.$$
Since $\rho<1$, iterating this recursion gives
$$A_n \leq \rho^n A_0+e^{18\theta}(1+\rho+\cdots+\rho^{n-1}) \leq A_0+\frac{e^{18\theta}}{1\!-\!\rho}<\infty.$$
Hence
\begin{equation}
    \sup_{n\ge0} A_n = \sup_{n\ge0}\mathbb E[V_{9n}] = \sup_{n\ge0}\mathbb E[e^{\theta |W_{9n}|}]<\infty. \label{eq:sup-W-9n}
\end{equation}
Now, let $m=9n+r$, where $0 \leq r < 9$.
During these $r$ progress epochs, the imbalance changes by at most $r$, and therefore
$$|W_m| \leq |W_{9n}|+8.$$
Thus,
$$e^{\theta |W_m|} \le e^{8\theta}e^{\theta |W_{9n}|}.$$
Taking expectations and using \Cref{eq:sup-W-9n},
$$\sup_{m\ge0}\mathbb E[e^{\theta |W_m|}]<\infty.$$
And, specifically,
$$\sup_{m\ge0}\mathbb E[\left|W_m\right|] = \sup_{m\ge 0}\mathbb E\!\left[|D_{\tau_m}|\right]<\infty.$$
\end{IEEEproof}

\ternaryproj*
\begin{IEEEproof}
For a word $\bfw\in\Sigma_3^L$, let $\pi(\bfw)$ be the binary subsequence obtained by deleting every occurrence of the symbol $2$. For all sufficiently large $L$, set 
$$
m_L\coloneqq\left\lfloor\frac{2L}{3}-L^{2/3}\right\rfloor
$$
and let $G_L$ be the event that both $\pi(\bfx)$ and $\pi(\bfy)$ have length at least $m_L$. Let $N_{\bfx}\coloneqq|\pi(\bfx)|$ and $N_{\bfy}\coloneqq|\pi(\bfy)|$. Each variable is distributed as $\operatorname{Bin}(L,2/3)$ and has mean $\mu=2L/3$. Hoeffding's inequality gives, for every $s>0$,
$$
\Pr(N_{\bfx}\leq\mu-s)\leq\exp\left(-\frac{2s^2}{L}\right).
$$
Substituting $s=L^{2/3}$ and using $m_L\leq\mu-L^{2/3}$ gives
$$
\Pr(N_{\bfx}<m_L)\leq\exp\left(-2L^{1/3}\right).
$$
The same bound holds for $N_{\bfy}$. Hence, the union bound gives
$$
1-\Pr(G_L)\leq2\exp\left(-2L^{1/3}\right)
=\exp\bigl(-\Omega(L^{1/3})\bigr).
$$

On $G_L$, let $\bfu$ and $\bfv$ be the length-$m_L$ prefixes of $\pi(\bfx)$ and $\pi(\bfy)$, respectively. Consider any ternary schedule of length $t$ for $(\bfx,\bfy)$ and delete all slots in which the emitted symbol is $2$. The remaining emitted symbols form the binary periodic sequence $(0,1,0,1,\ldots)$, and the remaining actions complete $(\bfu,\bfv)$. Among the first $t$ slots, at most $2t/3+1$ emit either $0$ or $1$. Applying this to an optimal ternary schedule gives
$$
T_{*}(\bfx,\bfy)\geq\frac{3}{2}\bigl(T_{*}(\bfu,\bfv)-1\bigr).
$$
The event $G_L$ depends only on the locations of the retained symbols. Conditional on $G_L$, the retained symbols are independent and uniform on $\{0,1\}$. Consequently, $(\bfu,\bfv)$ is a pair of independent uniformly distributed binary words of length $m_L$. Taking expectations gives
\begin{align*}
\mathbb{E}[T_{*}(\bfx,\bfy)]
&\geq \Pr(G_L)\frac{3}{2}
\left(\tau_2(m_L)-1\right)\\
&=\frac{3}{2}\tau_2(m_L)-O(1).
\end{align*}
The last equality follows from the exponential bound on $\Pr(G_L^c)$ and the deterministic estimate $\tau_2(m_L)\leq4m_L = O(L)$, thus the failure-event contribution is $O(e^{-\Omega(L^{1/3})}\cdot L) = o(1)$.
Under the assumed binary lower bound, the relation $m_L=2L/3-o(L)$ gives
$$
\tau_3(L)\geq\frac{3}{2}\left(2.03m_L-o(L)\right)-O(1)=2.03L-o(L).
$$
\end{IEEEproof}


\nearalternatinginterleavingbound*
\begin{IEEEproof}
For $j\geq0$, let
$$
a_j\coloneqq
\left|\left\{\boldsymbol{\omega}\in\{\mathrm{R},\mathrm{B}\}^{j}
\mathrel{\big|}\boldsymbol{\omega}\text{ has no factor }\mathrm{RRBB}\right\}\right|.
$$
Two occurrences of $\mathrm{RRBB}$ cannot overlap, because none of its nonempty proper prefixes is also a suffix.
Consequently, among the $2a_{j-1}$ one-symbol extensions of avoiding words, exactly $a_{j-4}$ create a terminal occurrence of $\mathrm{RRBB}$.
Thus,
$$
a_0,a_1,a_2,a_3=1,2,4,8,
\qquad
a_j=2a_{j-1}-a_{j-4}\quad(j\geq4),
$$
or, equivalently, let 
$$A(x) \coloneqq \sum_{j\geq0}a_jx^j = 1/Q(x),$$
where $Q(x) \coloneqq 1-2x+x^4$. We next decompose $A(x)$ to extract and bound its coefficients, i.e., to prove the uniform exponential bound $a_j \leq \kappa \lambda^j$.

Substituting $x=1/t$ into $Q(x)=0$ and multiplying by $t^4$ gives $t^4-2t^3+1=(t-1)(t^3-t^2-t-1)=0$, so, apart from the root $t=1$, its three roots are the roots of $t^3-t^2-t-1$.
Recall that $\lambda$ is the real root greater than $1$, and let $\mu,\bar{\mu}$ be the two nonreal roots.
It follows that
$$ Q(x)=(1-x)(1-\lambda x)(1-\mu x)(1-\bar{\mu}x).$$
For $t\in\{1,\lambda,\mu,\bar{\mu}\}$, write $Q(x)=(1-tx)R_t(x)$, and then $Q^{\prime}(x)=-tR_t(x)+(1-tx)R_t^{\prime}(x)$. At $x=1/t$, $Q^{\prime}(1/t)=-tR_t(1/t)$. Moreover, $Q^{\prime}(x)=-2+4x^3$, thus $Q^{\prime}(1/t)=-2+4t^{-3}$.
We conclude that the expression for the coefficient $C_t$ of $(1-tx)^{-1}$ in the partial-fraction decomposition of $1/Q(x)$ is therefore
$$C_t = \lim_{x\to1/t}\frac{1-tx}{Q(x)} = \frac{1}{R_t(1/t)} = -\frac{t}{Q^{\prime}(1/t)} = \frac{t}{2-4t^{-3}}.$$
In particular,
$$C_1=-\frac12, \quad C_\lambda=\kappa\coloneqq\frac{\lambda}{2-4\lambda^{-3}}, \quad C_\mu=B, \quad C_{\bar{\mu}}=\bar{B},$$
where $B\coloneqq\frac{\mu}{2-4\mu^{-3}}$.
It follows that
$$
\frac{1}{1-2x+x^4}
=
\frac{\kappa}{1-\lambda x}
-\frac{1/2}{1-x}
+\frac{B}{1-\mu x}
+\frac{\bar{B}}{1-\bar{\mu}x},
$$
Using $\frac{1}{1-tx} = \sum_{j \geq 0} t^jx^j$ and $A(x)=\sum_{t} \frac{C_t}{1-tx}$, we obtain
$$A(x)=\sum_{j \geq 0} \left( \sum_{t\in\{1,\lambda,\mu,\bar{\mu}\}} C_t t^j \right)x^j.$$
Therefore, a comparison of the coefficient of $x^j$ gives
$$a_j= \sum_{t\in\{1,\lambda,\mu,\bar{\mu}\}} C_t t^j.$$
which in this case yields 
$$ a_j = \kappa\lambda^j-\frac12 + B\mu^j + \bar{B} \bar{\mu}^{\,j}.
$$
By Vieta's formula, the product of the three roots of $t^3-t^2-t-1$ is $1$, so
$|\mu|=\lambda^{-1/2}$.
The estimates $1.839<\lambda<1.840$ imply by direct calculation that $ |\mu|<0.738$, and $|B|<0.092 $.
Consequently, for every $j\geq0$,
$$ a_j \leq \kappa \lambda^j - \frac12 + 2(0.092)(0.738)^j
\leq \kappa\lambda^j,$$
Since $0.738^j \leq 1$, we have $2(0.092)(0.738)^j \leq 0.184<\frac12$, and therefore $- \frac12 + 2(0.092)(0.738)^j \leq 0$.

Now fix $\bfz$ as in the lemma, and let $s$ be the number of its equal adjacent pairs.
Cutting $\bfz$ at these $s$ positions produces $s+1$ nonempty maximal alternating segments of lengths
$\ell_1,\ldots,\ell_{s+1}$, where $\sum_i\ell_i=2L$.
Every alternating length-four factor lies within one segment.
Hence, on the $i$-th segment, a reduced witness has no factor $\mathrm{RRBB}$, and there are at most $a_{\ell_i}$ possible colorings.
By \Cref{lemma:reduced-witness-bound},
$$ |D(\bfz)| \leq \prod_{i=1}^{s+1}a_{\ell_i} \leq \prod_{i=1}^{s+1}\kappa\lambda^{\ell_i} = \lambda^{2L}\kappa^{s+1}.$$
Finally, $s+1=2L-\rho_{\bfz}\leq2\beta L$, and $\kappa>1$, so
$$ |D(\bfz)|\leq\lambda^{2L}\kappa^{2\beta L}.$$
\end{IEEEproof}

\bibliographystyle{IEEEtran}
\bibliography{bibli}

\end{document}